\documentclass[sigconf,nonacm]{acmart}

\AtBeginDocument{%
  }

\usepackage{bm}
\usepackage{multirow}
\usepackage{subcaption}
\usepackage{url}
\usepackage{hyperref}
\usepackage{graphicx}
\usepackage{soul}
\usepackage{xcolor}
\sethlcolor{yellow!30}

\setcopyright{none}
\begin{document}

\title{More Isn't Always Better: Balancing Decision Accuracy and Conformity Pressures in Multi-AI Advice}

\author{Yuta Tsuchiya}
\email{ytsuchi28054@g.ecc.u-tokyo.ac.jp}
\affiliation{%
  \institution{The University of Tokyo}
  \city{Tokyo}
  \country{Japan}
}
\author{Yukino Baba}
\email{yukino-baba@g.ecc.u-tokyo.ac.jp}
\affiliation{%
  \institution{The University of Tokyo}
  \city{Tokyo}
  \country{Japan}
}


\begin{abstract}
  Just as people improve decision-making by consulting diverse human advisors, they can now also consult with multiple AI systems. Prior work on group decision-making shows that advice aggregation creates pressure to conform, leading to overreliance. However, the conditions under which multi-AI consultation improves or undermines human decision-making remain unclear. We conducted experiments with three tasks in which participants received advice from panels of AIs. We varied panel size, within-panel consensus, and the human-likeness of presentation. Accuracy improved for small panels relative to a single AI; larger panels yielded no gains. The level of within-panel consensus affected participants' reliance on AI advice: High consensus fostered overreliance; a single dissent reduced pressure to conform; wide disagreement created confusion and undermined appropriate reliance. Human-like presentations increased perceived usefulness and agency in certain tasks, without raising conformity pressure. These findings yield design implications for presenting multi-AI advice that preserve accuracy while mitigating conformity.
\end{abstract}



\maketitle

\section{Introduction}

Artificial Intelligence (AI) is increasingly used for decision support in domains such as medicine \cite{medi1, ex_medi, medi2}, law \cite{criminal, criminal2}, and education \cite{education}. In these applications, AI provides recommendations and humans make the final decision by evaluating them, which is expected to yield higher accuracy by leveraging complementary strengths \cite{collaboration, collaboration2}. Notably, human-AI decision-making research has primarily focused on settings involving a single AI advisor \cite{hadm_survey, XAI_ex, confidence, adult_ex}, whereas people often consult multiple sources of advice, such as seeking second opinions in medicine or relying on multiple peer reviews in academia \cite{multiple_advice, advice_survey}.

Analogously, people increasingly interact with multiple AIs. Research labs and companies continue to release new models, and more than 120,000 large language models (LLMs) are now registered on the Hugging Face library~\footnote{\url{https://huggingface.co/models}}. Because models are trained on different data and algorithms, they exhibit distinct strengths and biases\mbox{~\cite{harness}}. Reflecting this trend, multi-LLM workspaces such as TeamAI~\footnote{\url{https://teamai.com/multiple-models}} allow users to submit the same prompt to multiple LLMs in parallel and compare their outputs side-by-side for tasks like creative writing and coding.

Beyond such tools, multi-AI interaction is beginning to appear across a range of applied domains~\cite{finance,argumentative_experience, see_wide, multi_social_group, choicemates, postermate}. Examples include financial decision support~\cite{finance}, design and creativity support where users receive critical feedback from multiple AI personas~\cite{postermate}, and consumer decision-making informed by AI agents~\cite{choicemates}. Meanwhile, as AI-generated content becomes increasingly common in society, users have come to naturally encounter multiple AI-produced opinions and comments on social media and online platforms. Accordingly, recent research explores multi-AI systems that can support more reflective opinion formation and deliberation~\cite{argumentative_experience, multi_social_group, see_wide}.

This leads to a natural question: Is increasing the number of advisors always beneficial? Aggregating multiple perspectives has been shown to improve accuracy by correcting biases and surfacing new insights \cite{benefit_ad_op,ad_evi}. However, empirical findings suggest that the effects are more complex. Interestingly, simple statistical aggregation of multiple opinions (e.g., averaging or majority voting) usually outperforms humans’ deliberative selection of advice in terms of accuracy\mbox{~\cite{op_sel_ave}}. Conformity is a major reason for degraded integration: multiple advisors create pressure to follow the majority even when it is wrong\mbox{~\cite{asch, blackwell, asch2, conf_wisdom}}. Moreover, non-human agents can elicit conformity pressure differently from humans, depending on factors such as task uncertainty and agents’ appearance\mbox{~\cite{argumentative_experience, multi_social_group, dixit, robo_conf, robo_conf_3}}. These findings suggest that decision-support systems with multiple AIs should be carefully designed to manage conformity pressures and help users integrate advice more effectively.

\begin{figure*}[ht]
  \centering
  \includegraphics[width=\linewidth]{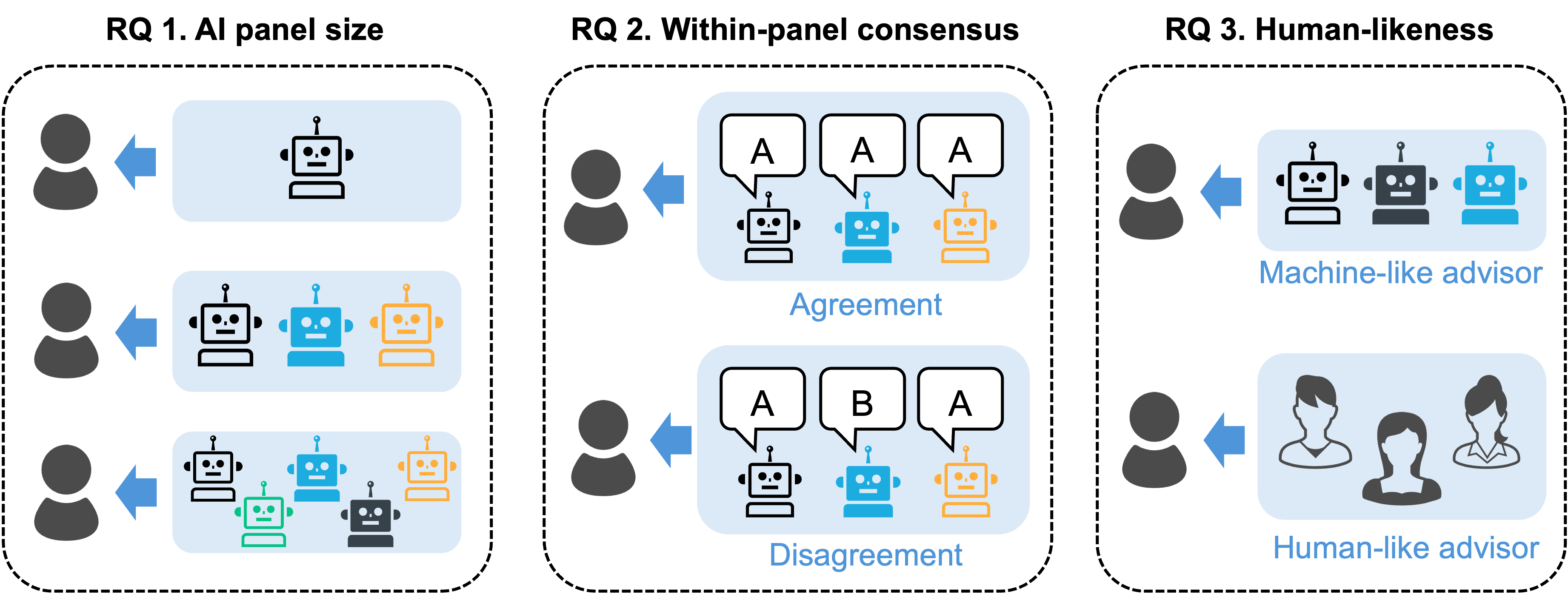}
  \caption{Overview of the experimental conditions corresponding to the three research questions (RQs). We varied (1) the number of AI advisors (RQ1), (2) the distribution of their opinions (RQ2), and (3) the presentation style of the agents (RQ3).}
  \Description{Experimental conditions visualized with icons. A diagram shows the three manipulations: panel size (1, 3, or 5 advisors), within-panel consensus versus disagreement, and advisor appearance as either robot or human faces.}
  \label{RQ_2}
\end{figure*}

However, these issues have received limited attention in prior research on human-AI decision-making~\cite{more_advice, human_ai_conf, AI_human_comb,ensemble}. For example, second-opinion studies examine how additional advice influences reliance on an initially presented AI recommendation, without directly addressing conformity pressures~\cite{more_advice}. Other studies of human-AI teams investigate how people differentially weight advice from humans versus AI systems \cite{human_ai_conf, AI_human_comb}, but do not consider interactions among multiple AI advisors. Consequently, it remains unclear whether presenting panels of multiple AIs improves decision accuracy by increasing informational diversity or heightens conformity pressure that leads to incorrect decisions. Addressing this gap is essential for the principled design of \textit{Human-Multi-AI Decision-Making}.

In this paper, we aim to clarify how advice from multiple AI panels influences human decision-making. Focusing on three design factors directly linked to conformity pressure, we pose the following three research questions (Figure\mbox{~\ref{RQ_2}}):

\paragraph{\textbf{RQ1: How does the AI panel size shape human decision-making?}} Increasing the number of human advisors can broaden informational diversity; however, larger groups can heighten conformity pressure and cognitive load, which may in turn reduce decision accuracy\mbox{~\cite{advice_survey}}. Prior work also suggests that larger AI panels can trigger resistance or polarization, especially in discussions of value-sensitive social issues~\cite{multi_social_group}. By contrast, in accuracy-critical decision contexts, panel consensus may instead be interpreted as stronger collective evidence and lead to greater conformity pressure~\cite{epistemic}. Thus, AI panel size can have competing effects on decision-making and should be carefully designed.

\paragraph{\textbf{RQ2: How does the within-panel consensus of AIs affect human decision-making?}} High consensus in human groups is known to strengthen conformity pressure\mbox{~\cite{asch}}. Minority dissent, by contrast, can promote deliberation but may also be ignored or cause confusion\mbox{~\cite{multiple_advice, agregate_source}}. Social-psychology work on majority influence repeatedly shows that both overall group size and majority size are key determinants of conformity\mbox{~\cite{blackwell}}. However, it remains unclear how such consensus of AIs affects decision accuracy.

\paragraph{\textbf{RQ3: How does the human-likeness of AI panels influence human decision-making?}} Prior work suggests that increasing anthropomorphism can strengthen the social influence of non-human agents, heightening users' willingness to rely on their recommendations~\cite{human_like_conf,anth_llm,del_anth,robo_social}. Such human-likeness cues may have a stronger impact on conformity than the number of agents~\cite{robo_att_size}. LLMs are increasingly perceived as social partners, and cues such as tone and persona can shape how users engage with AI advice~\cite{tone_aware, AIlove, persona, llm_personal, llm_personal2}. When presented in panels, human-like AIs may be perceived as a social group that elicits stronger conformity pressure. Thus, perceived human-likeness is an important factor in understanding how multi-AI systems may undermine decision accuracy.

\vskip\baselineskip

To address the research questions above, we conducted an online experiment via crowdsourcing with 348 participants. 
In Study 1, we manipulated the number of AI panels (1, 3, or 5) and observed the resulting within-panel consensus. In Study 2, we fixed the panel size at three AIs and manipulated the degree of anthropomorphism in their presentation. The main findings of this study are as follows:

\begin{itemize}
\item \textbf{AI panel size}: Small panels improved accuracy compared to a single AI, but larger panels yielded no additional benefit. Increasing panel size did not strengthen reliance linearly, suggesting the existence of an optimal panel size.
\item \textbf{Within-panel consensus}: High consensus fostered overreliance, while a single dissent reduced conformity pressure. Wide disagreement, however, led to confusion due to information overload, undermining appropriate reliance.
\item \textbf{Human-likeness}: Human-like presentation increased perceived usefulness and agency in certain tasks. Conformity pressure showed individual differences, but did not affect performance on average.
\end{itemize}

These results highlight the need for design principles that preserve decision accuracy by supporting appropriate reliance while managing conformity pressures.

\section{Related Work}

\subsection{Human-AI decision-making}

Recent research in HCI and AI has examined frameworks for collaborative decision-making with AI systems~\cite{collaboration2, hadm_survey}. Although humans are expected to complement AI by correcting its errors and biases, collaborative performance is often suboptimal because of inappropriate reliance on AI~\cite{collaboration, reliance, accuracy}. To address this, some work has aimed to calibrate trust by providing explanations for AI recommendations~\cite{ex_medi, counterfactual, XAI_ex}, while other studies have focused on adjusting users' self-confidence~\cite{confidence, know_ab_know, confidence_align}. Yet, a general approach that reliably improves decision accuracy with AI assistance has not been established.

The advent of LLMs has transformed decision support by enabling interactive, language-based collaboration~\cite{shap, LLM_XAI, text_trust, contrastive}. Deliberative systems using dialog-based interaction through LLMs have been shown to promote appropriate reliance~\cite{deliberation}, and adaptive frameworks have been shown to selectively present salient features expected to have beneficial effects on human decision-making~\cite{text_trust}. However, the confident and persuasive style of LLM outputs may foster overreliance, prompting recent work on interventions to mitigate this risk~\cite{LLM_reliance, to_use_or}.

On the other hand, most existing empirical studies have assumed scenarios in which a single AI provides advice to humans. Decision-making scenarios involving multiple AIs have received limited attention; most existing approaches focus on human-AI comparisons rather than potential conformity effects by multiple advisors \cite{more_advice, AI_human_comb}. Other work on ensembles of multiple machine learning models has examined how reliance differs between ensemble and single-model predictions\mbox{~\cite{ensemble}}, and proposed explanation methods for the ensemble prediction process\mbox{~\cite{ensemble_int}}. However, these works view the ensemble as a single predictive entity, whereas we study multiple AIs presented as independent advisors. Consequently, little is known about how conformity pressures by multiple AI advisors shape individual decision-making.

\subsection{Using advice from multiple sources}

Humans have long sought to improve decision accuracy by incorporating multiple sources of advice. Experimental studies have demonstrated that introducing multiple advisors can enhance both the accuracy and quality of decisions across a variety of contexts~\cite{multiple_advice, advice_survey, agg_ex_op}. Several theoretical and empirical mechanisms underlie these benefits. First, error cancellation: when combining multiple independent opinions, the random errors in each judgment tend to offset one another, bringing the average estimate closer to the true value~\cite{ave_judge, condorcet, ad_evi, ad_evi2}. Second, information diversity: aggregating advice from sources that draw on different perspectives or information helps correct individual biases and enrich the informational basis of decisions~\cite{multi_random, agg_ex_op, multi_random2}. The effect is particularly pronounced when the advice is highly independent or uncorrelated~\cite{multi_random2}.

However, the presentation of multiple opinions is not invariably beneficial. People prefer to evaluate and select opinions, but rarely beat simple aggregation such as averaging\mbox{~\cite{op_sel_ave}}. Decision-makers often overvalue their own estimates while undervaluing others' advice~\cite{multiple_advice}. Disagreement among advisors can sometimes improve accuracy, but it may also reduce confidence, weaken trust, and create confusion that harms performance~\cite{multi_chaos, robo_chaos2, advice_survey}. Moreover, social influence, such as conformity and groupthink, can weaken critical judgment and let majority advice dominate decisions~\cite{asch, groupthink}. Thus, multiple sources of advice offer both potential benefits and risks: it is key to clarify how multi-AI panels shape these outcomes.

\subsection{Conformity pressures}

Conformity refers to the phenomenon in which individuals adjust their behavior or judgments to align with those of others~\cite{conformity_is}. It is generally understood to consist of two forms: informational conformity, in which people assume that others are more likely to be correct, and normative conformity, in which people align with the group to avoid social costs and gain acceptance~\cite{conf_neuro, conformity_kind}.
A landmark demonstration of this phenomenon is Asch's classic study~\cite{asch}. Even in a perceptually unambiguous task, when all surrounding peers provided the same incorrect response, participants conformed on average in 37\% of the trials.

Prior work offers mixed views on whether non-human agents can generate conformity pressure. The CASA (Computers Are Social Actors) paradigm shows that people mindlessly apply social rules and expectations to computers, and has expanded from robots and chatbots to recent generative AI\mbox{~\cite{casa,casa_deep,CASA_media,CASA_genAI}}. A large body of empirical work supports this view, while also suggesting that conformity dynamics with non-human agents can partly differ from those in human groups\mbox{~\cite{robo_conf_base, robo_conf_base2}}. For example, in simple perceptual tasks, people are less likely to conform to robots or virtual avatars than to humans; yet in high-uncertainty tasks such as emotion estimation, informational conformity toward robots becomes stronger\mbox{~\cite{robo_conf, robo_conf_task, robo_conf_2, robo_conf_3}}. Although normative conformity is generally weaker, it can increase when robots' appearance or behavior is made more human-like\mbox{~\cite{human_like_conf,robo_conf_4}}.

Recent LLM studies likewise report normative conformity driven by a sense of belonging: multiple LLM personas sharing participants' identities can promote prosocial behaviors such as donating, or shift confirmation bias on social issues~\cite{multi_norms,multi_social_group}.
One study suggests that informational conformity may not arise; users treat AI as a mere computation or search tool and do not interpret multi-agent agreement as increased reliability~\cite{multi_social_group}. However, it examines opinion change on societal issues without ground truth, where participants may prioritize alignment with perceived norms over assessing correctness.
In contrast, our study targets decisions where accuracy matters. As prior robot/virtual-agent research and CASA theory suggest, people may perceive AI as a human-like expert advisor~\cite{casa,epistemic}. Therefore, it is important to empirically test whether consensus among multiple AI agents can elicit informational conformity, and to examine how this affects decision accuracy.

\section{Experimental Design}

We designed binary prediction tasks in which participants made judgments based on AI's advice. In Study 1, the AI panel size was manipulated to test RQ1 between participants, with within-participant variation used for RQ2. In Study 2, we manipulated the human-likeness of AI presentation and compared it with three-AI panels from Study 1 to examine RQ3. This study was reviewed and approved by the Ethical Review Committee for Experimental Research involving Human Subjects at the University of Tokyo.

\subsection{Decision-making tasks}

We employed three types of tasks: Income, Recidivism, and Dating prediction~\cite{adult_2, ofer2013compas, speeddating1}. These tasks are widely used in human-AI decision-making research and cover a broad range of contexts with distinct social and cognitive dimensions~\cite{hadm_survey}.
Our goal is to examine how people integrate AI advice under uncertainty, so we selected tasks calibrated to yield 60--70\% accuracy in the human-only condition, making them understandable yet challenging\mbox{~\cite{more_advice, text_trust}}. Income and Recidivism prediction are not everyday tasks. However, prior work suggests that conformity can be stronger in unfamiliar domain, so testing advice-taking under such conditions is important\mbox{~\cite{conf_cmbiguity, robo_conf_4}}. Including a relatively familiar task (Dating) further widens the scope and strengthens generalizability.

For each task, we sampled 50 test cases and 500 training cases while preserving the original label distribution of the datasets. Prior to analysis, missing values were removed during preprocessing. The feature sets were adjusted with reference to prior literature to minimize differences in the number of features across tasks:
\begin{itemize}
\item \textbf{Income}~\cite{adult_ex, adult_2, Yang_etal}: Participants were asked to predict whether an individual's annual income exceeded or fell below USD 50,000 based on a provided profile. Each profile included seven features: gender, age, education level, marital status, occupation, type of work, and weekly working hours. The profiles were generated from the UCI Adult Dataset~\cite{adult_2}, which is based on the U.S. Census, and participants were informed that the profiles reflected standards in the United States around 1994.

\item \textbf{Recidivism}~\cite{ofer2013compas, tone_aware, two_heads}: Participants predicted whether a defendant would reoffend within two years, based on profiles from the COMPAS dataset (Broward County, Florida, 2013--2014)~\cite{ofer2013compas}. Each profile contained eight features: demographic information (gender, age, race), criminal history (number of prior non-juvenile offenses, juvenile misdemeanors, and juvenile felonies), as well as current charge and charge severity. As recidivism prediction is framed as a high-stakes judicial task, which is widely recognized as contributing to underreliance on AI advice~\cite{aversion_high}, participants were reminded of its real-world significance to ensure thoughtful engagement.

\item \textbf{Dating}~\cite{speed_date_data, speeddating1}: Participants saw profiles from a speed-dating dataset (2002--2004)~\cite{speed_date_data}, including an individual's demographics (gender, age, race), a partner's demographics (age, race), and impression ratings (attractiveness, sincerity, intelligence, fun, ambition, shared interests). Based on these 11 features, they predicted in binary form whether the focal individual would want to meet their partner again. Note that predictive tasks grounded in subjective judgments often elicit algorithm aversion, where people tend to underweight or ignore algorithmic advice~\cite{alg_aversion}.

\end{itemize}

\subsection{Interface}

We employed the Judge-Advisor System (JAS) paradigm to design the decision-making tasks~\cite{JAS}. JAS is a classic framework that has been widely adopted in human-AI decision-making research~\cite{collaboration2, confidence}. The paradigm consists of three stages: (1) an initial human prediction, (2) AI-provided advice, and (3) the human's final decision. By comparing the initial and final judgments, we were able to compute multiple measures of AI reliance.

\begin{figure*}[ht]
  \centering
  \includegraphics[width=\linewidth]{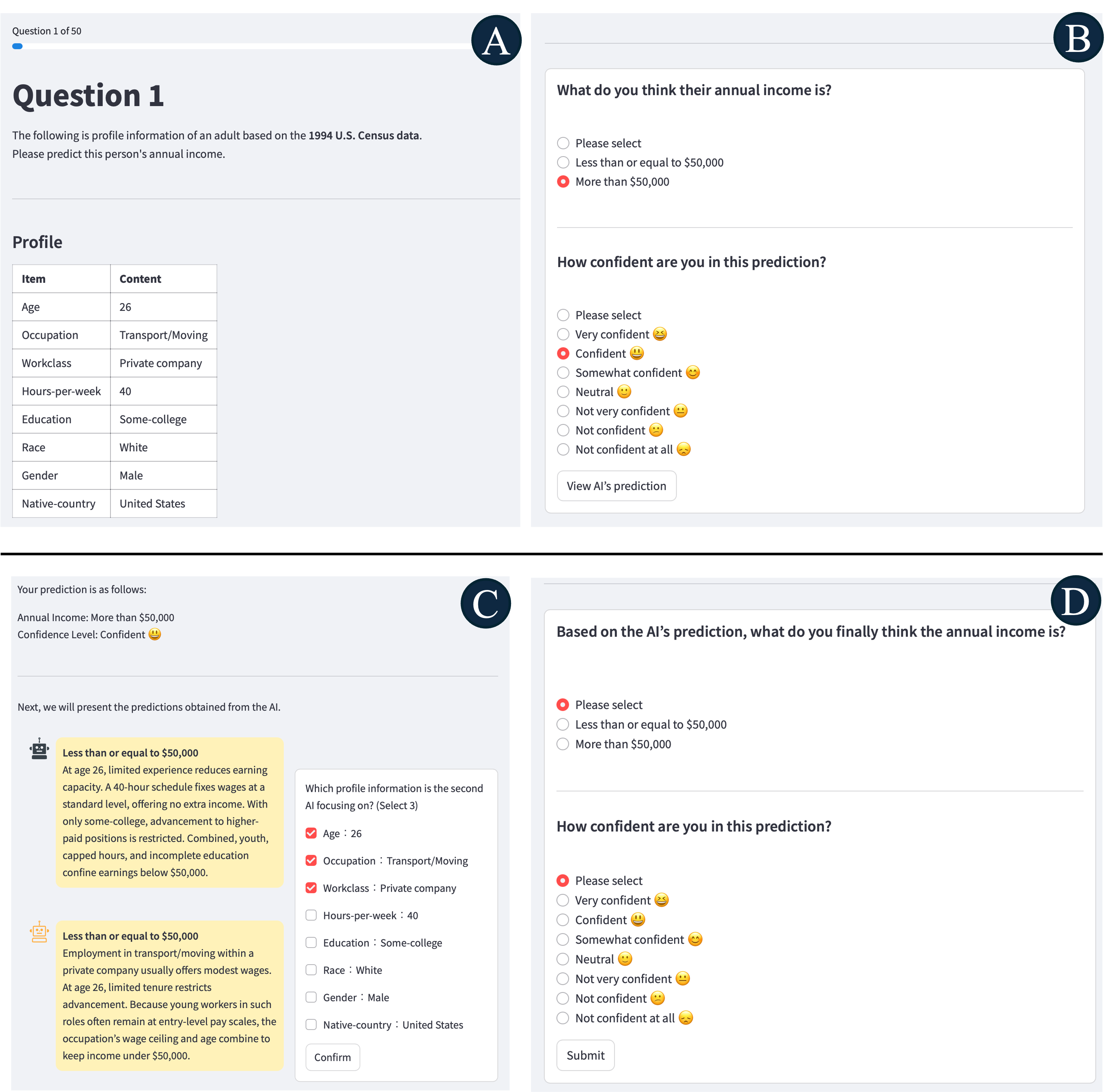}
  \caption{User interface for a single trial in the Income task. (A) Problem presentation. (B) Initial prediction. (C) AI advice with explanation and attention check (shown sequentially for multiple AIs). (D) Final decision. For (C), the figure illustrates the attention check of the second AI’s prediction.}
  \Description{Interface for one trial in the Income task. Four panels show the workflow: (A) participant profile display, (B) initial prediction and confidence rating, (C) AI advice with explanation and feature check, and (D) final prediction with updated confidence.}
  \label{interface1}
\end{figure*}

As shown in Figure~\ref{interface1}, the specific flow of a single trial in this experiment is as follows.

 \begin{enumerate}
    \item[(A)] \textbf{Problem presentation:} Participants were presented with profile information about a target individual.
    \item[(B)] \textbf{Initial prediction:} Based on the information, participants made a binary prediction and reported their confidence on a 7-point Likert scale.
    \item[(C)] \textbf{AI advice and attention check:} Participants were shown the AI's prediction together with a natural language explanation. To ensure careful reading, participants were also asked to select the features highlighted by the AI. In conditions with multiple AIs, advice was presented sequentially with repeated checks.
    \item[(D)] \textbf{Final decision:} Participants made their final prediction and reported their confidence again.
\end{enumerate}

\subsection{AI model}

We simulated decision-making scenarios by presenting participants with individual decision trees sampled from a random forest as AI advisors~\cite{random_forest}. Because random forests generate numerous distinct trees through randomized sampling of data and features, natural variation arises in both predictions and rationales across models.
The AI models were calibrated to 70\% accuracy, reflecting agents that are imperfect yet reasonably reliable, as in prior studies~\cite{more_advice, text_trust}.
To keep performance comparable across single- and multi-AI conditions, we drew a diverse subset of models from a Rashomon set~\cite{rashomon} with similar average accuracy yet varied predictions at the individual-case level.
Concretely, we trained a random forest~\cite{random_forest} on 500 training cases and evaluated it on 50 test cases. Among the obtained models, we identified 30--36 trees that achieved exactly 70\% accuracy (35 correct out of 50 test cases) and adopted them as the Rashomon set. This procedure allowed us to simulate realistic collaborative decision-making contexts, in which multiple AIs present different perspectives and rationales.

During the experiments, a fixed number of trees were randomly sampled from this set for each participant and condition, and applied consistently across all task samples. For each task instance, we recorded the AI predictions, their rationales, the majority-vote label, as well as the input features and ground-truth label, which were subsequently used in the analysis.

\subsection{Explanation generation}

Explanations of the rationale behind AI predictions help users interpret model outputs and compare them with their own judgments~\cite{hadm_survey, XAI_ex}. 
In this study, we used LLM-generated natural-language explanations which can make AI reasoning easier to understand without requiring technical expertise~\cite{to_use_or}.
These explanations were generated by incorporating feature attribution information into the LLM prompt. Specifically, for each decision tree in the Rashomon set, we applied SHAP~\cite{shap} to extract the top three features contributing to its prediction. Then, we converted them into concise natural language explanations using GPT-4o via the OpenAI API~\footnote{\url{https//openai.com/blog/openai-api}}. Importantly, the LLM was instructed to supplement the SHAP-based rationale rather than replace it, which helped enhance interpretability while reducing the risk of hallucination. Such LLM-based explanation approaches have increasingly been adopted in recent work~\cite{deliberation, text_trust}.

\begin{figure}[htbp]
  \centering
  \includegraphics[width=\linewidth]{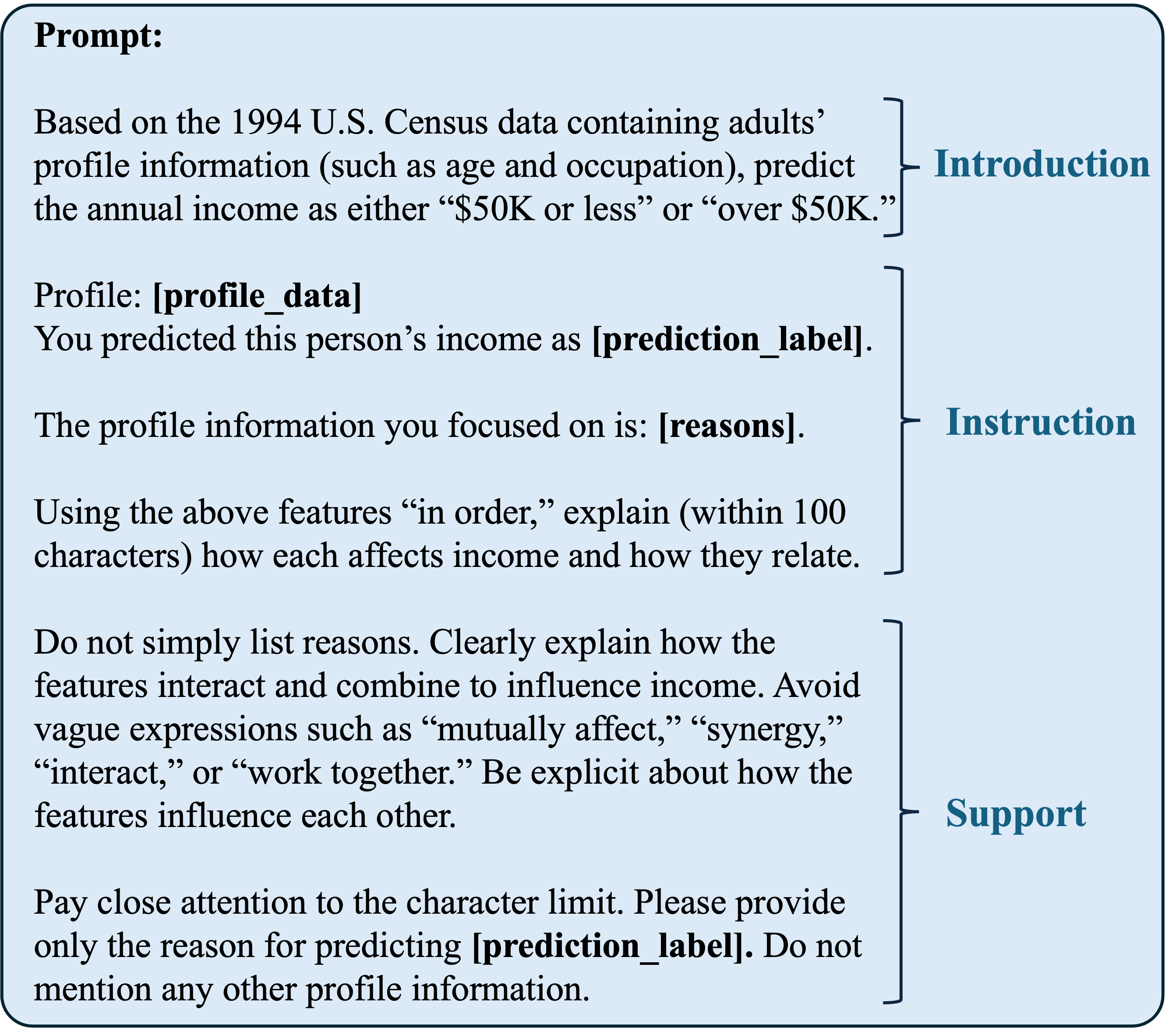}
  \caption{Prompt design for GPT-4o. The design consists of three main components: (1) introduction prompt, (2) instruction prompt, and (3) support prompt. Quoted text indicates variables that are replaced in each trial.}
  \Description{Structure of the GPT-4o prompt. Three parts: task introduction, input features with prediction and explanation, and supplementary clarifications and constraints.}
  \label{prompt}
\end{figure}

The prompt design for GPT-4o is illustrated in Figure~\ref{prompt}. It consisted of three main components: (1) an introduction prompt describing the target decision-making task; (2) an instruction prompt presenting the features, predicted label, and SHAP-based rationale, with a request to explain how these factors contributed to the prediction; and (3) a support prompt specifying additional constraints to concretize the explanation and minimize hallucination.

Participants were presented with each model's predicted label together with the generated explanation as shown in Figure~\ref{interface1}. In Study 2 (RQ3), we further modified the prompt by instructing changes in the tone of explanations.

\subsection{Participants}

Participants were recruited through the crowdsourcing platform Lancers~\footnote{\url{https://www.lancers.jp/}}. As inclusion criteria, Participants were required to be adults aged 18 or older, with an approval rate of $\geq 95\%$ based on their track record on the platform. Each participant was allowed to take part only once, with duplicate participation prevented using worker IDs. All tasks were conducted in Japanese, targeting residents of the Asian region. The use of generative AI during the experiment was explicitly prohibited.

A total of 365 individuals initially consented and completed the tasks. To ensure data quality, we embedded three attention-check items throughout the experiment, which required participants to select specific response options~\cite{caution_test}. Participants were excluded if they (i) failed three attention checks and completed the tasks in less than the 5th percentile of total completion time (indicating excessively short duration), or (ii) showed mechanical response patterns, such as selecting the same option across multiple items. Applying these criteria yielded a final sample of 348 participants. Study 1 recruited 260 participants, and Study 2 recruited an additional 88 participants.

We collected participants' demographic information based on their public profile data on Lancers. Across all conditions, the mean age was 44.5 years ($SD = 10.7$). The sample consisted of 58\% male and 42\% female. We also measured participants' frequency of AI use on a 7-point Likert scale (0 = never, 6 = very often). The ($M \pm SD$) score was 3.43 $\pm$ 1.55, indicating that participants used AI at a moderate frequency on average. No significant differences in AI-use frequency were observed across conditions.

Based on pilot testing, the estimated average completion time was approximately 50 minutes. Participants who completed the study were compensated USD 5.50, consistent with local minimum wage standards (USD 6--7/hour). They could withdraw at any time; however, because the tasks were administered as a single continuous session, they were completed without breaks.

\subsection{Procedure}

\begin{figure*}[ht]
  \centering
  \includegraphics[width=\linewidth]{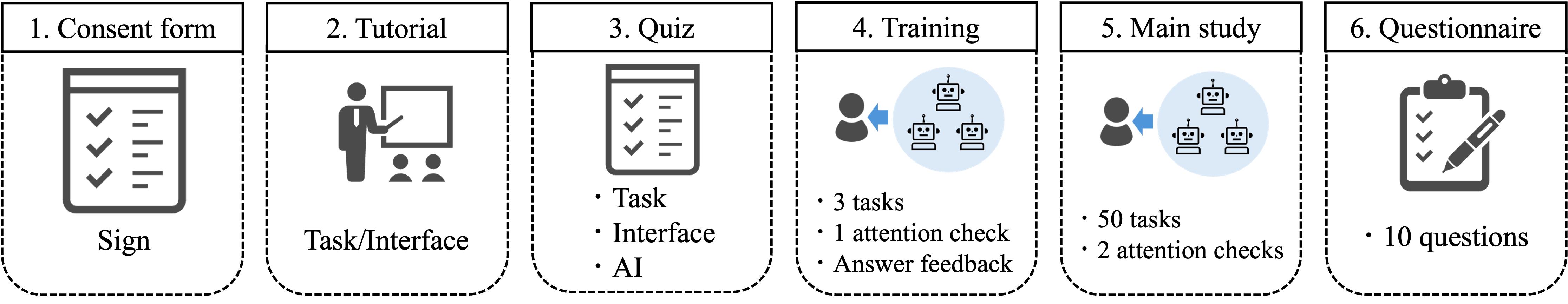}
  \caption{Overall procedure of the experiment. Participants were randomly assigned to one of the experimental conditions in Study 1 or Study 2. The study consisted of six sequential steps: (1) providing informed consent, (2) completing a tutorial on the interface and task, (3) answering a comprehension quiz, (4) performing three training tasks with immediate feedback, (5) completing 50 main decision-making tasks under the assigned condition, and (6) responding to a post-task questionnaire.}
  \Description{Experimental workflow shown as six sequential icons. A horizontal diagram depicts the study procedure as icons arranged in order: consent, tutorial, quiz, training, main study, and questionnaire.}
  \label{procedure}
\end{figure*}

The overall procedure of the experiment is illustrated in Figure~\ref{procedure}. Participants were randomly assigned to one of the experimental conditions in Study 1 or Study 2.

\begin{enumerate}
    \item \textbf{Consent form:} Participants read the study purpose, risks, and rights, and consented to the use of anonymized responses.
    \item \textbf{Tutorial:} The interface and task flow were introduced.
    \item \textbf{Quiz:} A short quiz was used to check participants’ understanding of the task (e.g., ``Can you change your prediction after AI advice?''). Only those with full scores advanced.
    \item \textbf{Training:} Three practice tasks allowed participants to learn the procedure, with correct answers revealed afterward. In addition, participants completed one attention check.
    \item \textbf{Main study:} Participants completed 50 prediction tasks with AI advice. They also completed two attention checks.
    \item \textbf{Questionnaire:} A post-task survey measured reliance, perceived conformity pressure, and cognitive load.
\end{enumerate}

\subsection{Measurement}

To capture the influence of multi-AI advice on human decision-making, we evaluated four dimensions: decision accuracy, reliance, confidence change, and subjective perceptions of AI.

\subsubsection{Decision accuracy}

To measure participants' performance, we calculated the proportion of correct final responses across the 50 test items. For RQ2, we also examined the accuracy of participants' initial predictions, allowing us to analyze how opinion diversity and consensus among AIs improved decision outcomes.

\begin{table*}[ht]
  \centering
  \caption{Questionnaire items.}
  \label{questionnaire_items}
  \begin{tabular}{l|l|l}
    \toprule
    \textbf{Category} & \textbf{No.} & \textbf{Question} \\
    \midrule
    \multirow{6}{*}{Reliance \& Conformity} & Q1 & Did you feel that the AI guided your opinion? \\
    & Q2 & Did you feel to change your thinking or follow the AI's hints? \\
    & Q3 & Did you feel any pressure from the AI, or did it come across as pushy? \\
    & Q4 & Did you feel that you made the decision on your own initiative? \\
    & Q5 & Do you think the responsibility for the decision's outcome lies with the AI? \\
    & Q6 & Did you feel that the AI was more intelligent than you? \\
    \midrule
    \multirow{4}{*}{AI Usefulness} & Q7 & Did the AI support confuse you?\\
    & Q8 & Did the AI support help you organize your thoughts? \\
    & Q9 & Did the AI support help reduce the burden of decision-making? \\
    & Q10 & Do you think the AI could help you complete tasks with less effort?\\

    \bottomrule
  \end{tabular}
\end{table*}

\subsubsection{Reliance}

To quantify reliance on AI advice, we adopted two widely used indicators:

\begin{itemize}
\item \textbf{Agreement Fraction}: the proportion of final responses that matched the AI's answer.
\item \textbf{Switch Fraction}: the proportion of cases where participants changed their initial answer to match the AI's answer.
\end{itemize}

We further evaluated reliance appropriateness when the participant's initial answer disagreed with the AI~\cite{know_ab_know}:

\begin{itemize}
\item \textbf{Relative Positive AI Reliance (RAIR):} the proportion of cases where participants switched to the correct AI's answer.
\item \textbf{Relative Positive Self-Reliance (RSR):} the proportion of cases where participants kept their own correct answer against an incorrect AI's answer.
\item \textbf{Accuracy with Initial Disagreement (Accuracy-wid):} the proportion of correct final answers when human and AI initially disagreed.
\end{itemize}

For conditions involving multiple AIs, disagreement among models was possible. In such cases, we defined \textit{reliance} as following the majority recommendation of the AI panel. This definition allows us capture conformity pressure driven by majority consensus.

\subsubsection{Confidence change}

In each task, participants reported their confidence level on a 7-point Likert scale (0 = not at all confident, 6 = very confident) along with both their initial and final predictions. This difference allows us to evaluate how the AI's advice affected changes in confidence.

\subsubsection{Subjective perceptions of AI}

We administered 10 questionnaire items using a 7-point Likert scale (0 = strongly disagree, 6 = strongly agree), supplemented by open-ended responses. The items covered: (1) Reliance and conformity tendencies (Q1--Q6)~\cite{contrastive, dixit}, and (2) Perceived usefulness of AI support (Q7--Q10)~\cite{deliberation, usefulness}.
The full set of questions is presented in Table~\ref{questionnaire_items}. These subjective evaluations, combined with behavioral indicators, enabled us to examine how participants perceived advice from multiple AIs.

\subsection{Analysis methods}

We selected statistical tests based on each measure's distribution. We checked normality (Shapiro-Wilk) and homogeneity of variance (Levene) for all dependent variables. Between-subject comparisons used independent t-tests or one-way ANOVA when assumptions held, and Mann-Whitney U or Kruskal-Wallis tests otherwise, with Tukey HSD or Dunn post hoc tests as appropriate. Within-subject pre-post comparisons (RQ2 and confidence changes) used paired t-tests or Wilcoxon signed-rank tests. For the three-level within-subject comparison in RQ2 (five AIs), we used repeated-measures ANOVA or Friedman tests, followed by Bonferroni-corrected pairwise t-tests or Wilcoxon tests. All p-values are reported for two-tailed tests, with significance at $p$ < .05.

\begin{figure*}[ht]
  \centering
  \begin{subfigure}[b]{0.33\linewidth}
    \centering
    \includegraphics[width=\linewidth]{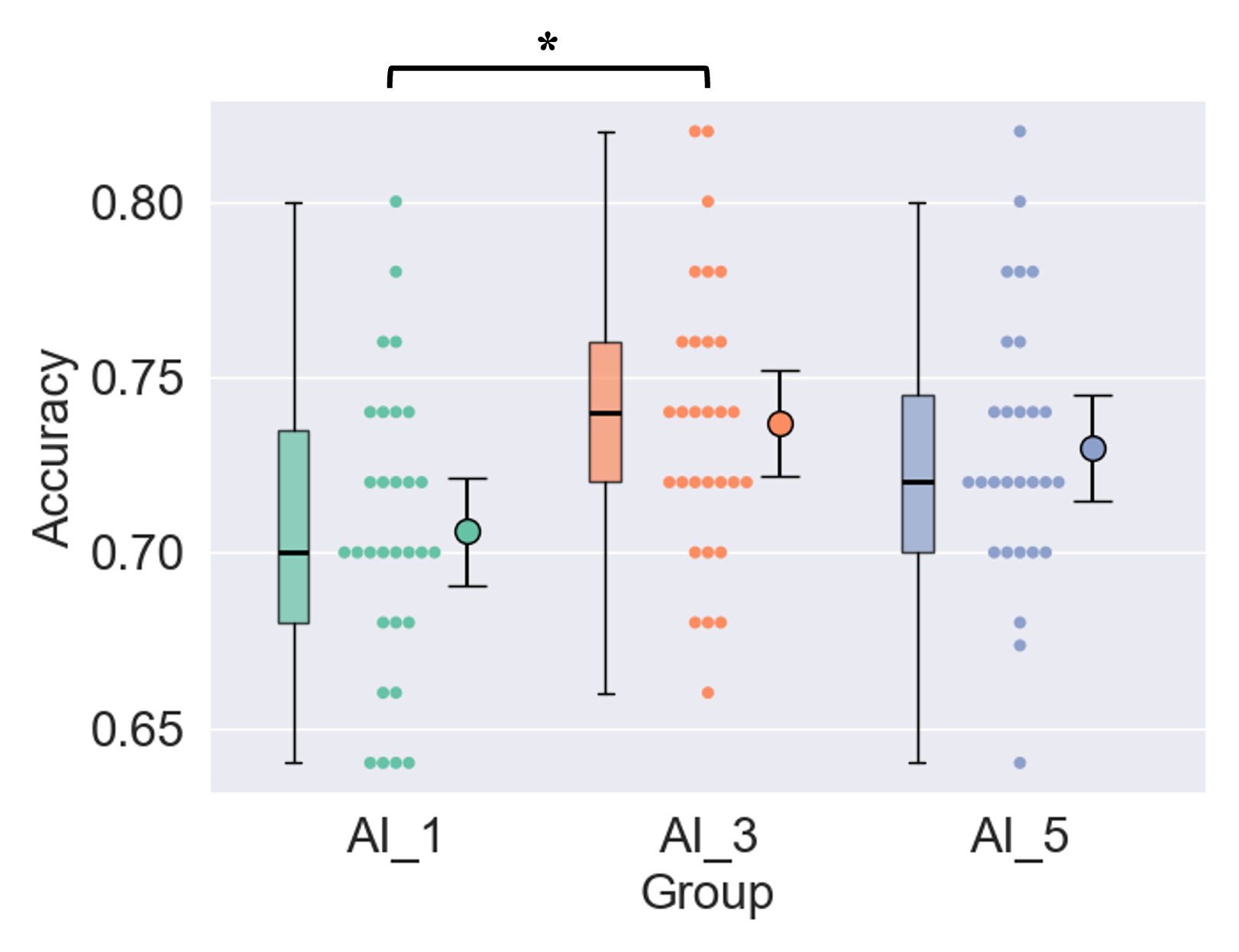}
    \caption{Income}
    \label{accuracy_AInum_a}
  \end{subfigure}
  \begin{subfigure}[b]{0.33\linewidth}
    \centering
    \includegraphics[width=\linewidth]{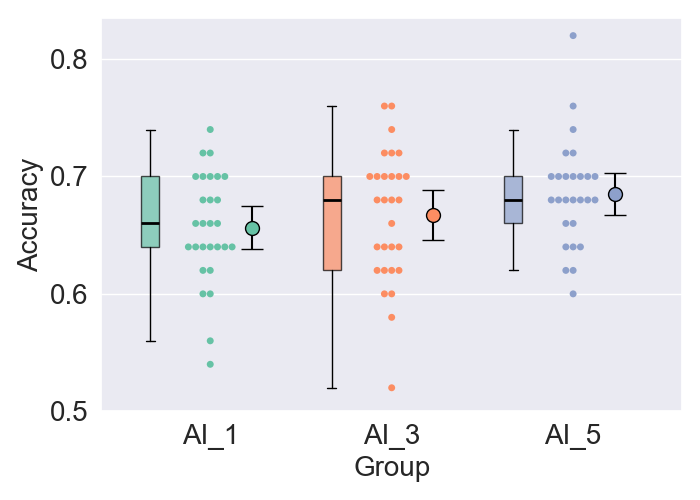}
    \caption{Recidivism}
    \label{accuracy_AInum_b}
  \end{subfigure}
  \begin{subfigure}[b]{0.33\linewidth}
    \centering
    \includegraphics[width=\linewidth]{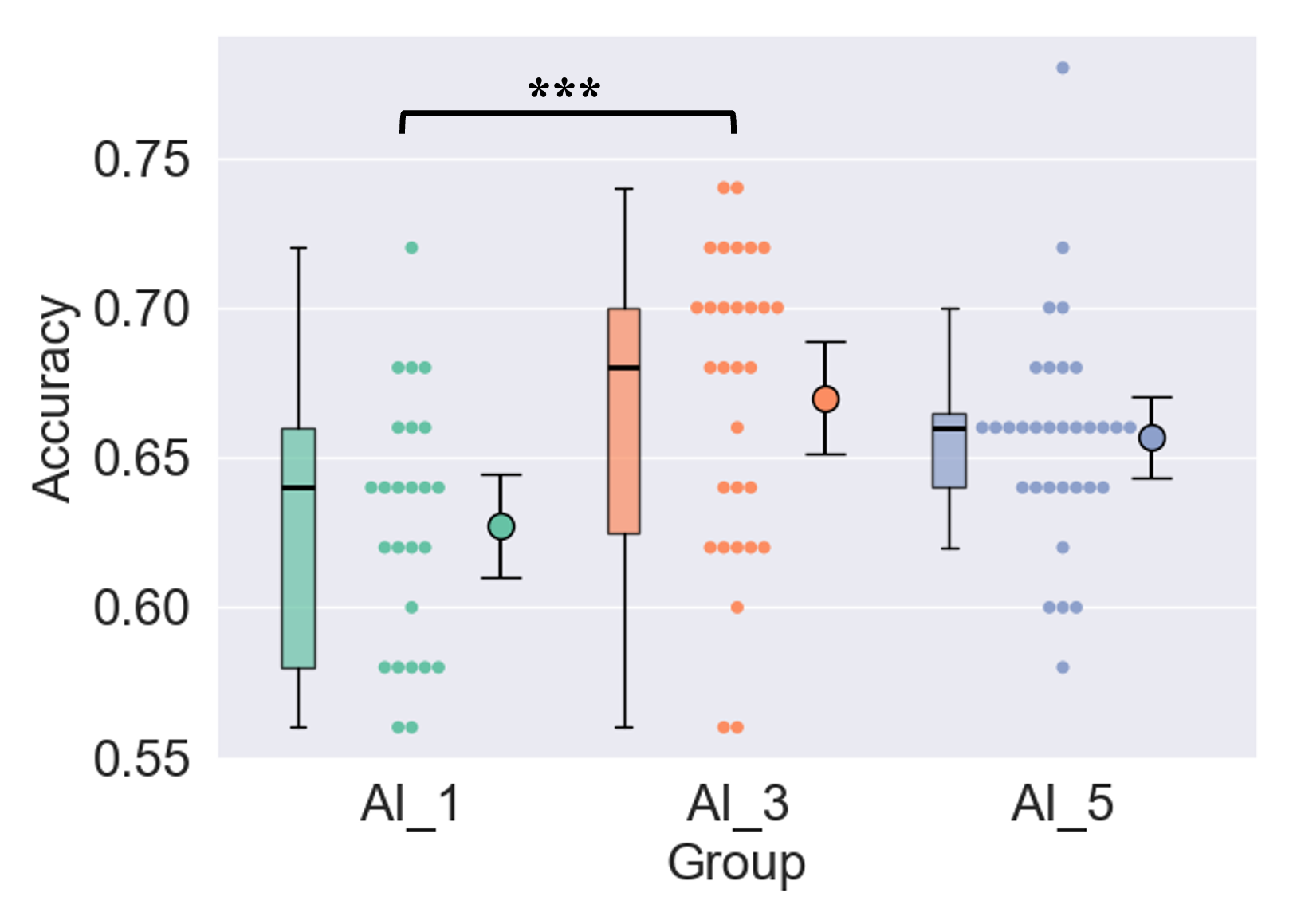}
    \caption{Dating}
    \label{accuracy_AInum_c}
  \end{subfigure}
  \caption{Comparison of decision accuracy across AI panel sizes. For each task, boxplots show the distribution of final prediction accuracy, with overlaid individual data points and mean values with 95\% confidence intervals. Significant differences between conditions are indicated (*: $p < .05$, ***: $p < .001$).}
  \Description{Decision accuracy by AI panel size across three tasks. Boxplots show accuracy for 1, 3, and 5 AI advisors in Income, Recidivism, and Dating tasks. Distributions and individual scores are plotted, with means and 95\% confidence intervals overlaid. Significant differences are marked above panels. Accuracy gains were most consistent with 3 AIs in Income and Dating tasks, while 5 AIs showed no further improvement.}
  \label{accuracy_AInum}
\end{figure*}

\section{Study 1}

\subsection{Setup}

We manipulated the number of AI panels (1, 3, or 5) as a between-subjects factor. Participants were randomly assigned to one of the conditions, and each participant completed all tasks under the panel size assigned to them. The allocation of participants to each group is summarized in Table~\ref{participants_study1}.

\begin{table}[t]
  \centering
  \caption{The number of participants for each task and AI panel size in Study 1.}
  \label{participants_study1}
  \begin{tabular}{l|ccc}
    \toprule
    \textbf{Task} & \textbf{Panel size: 1} & \textbf{Panel size: 3} & \textbf{Panel size: 5} \\
    \midrule
    Income       & 30 & 30 & 28 \\
    Recidivism   & 28 & 29 & 27 \\
    Dating       & 26 & 30 & 32 \\
    \bottomrule
  \end{tabular}
\end{table}

For participants assigned to the three- and five-AI conditions, we further categorized the distribution of opinions across the 50 tasks. Since the Rashomon set naturally generated diverse predictions through random sampling, these opinion distributions were treated as observed within-subject factors in the analysis for RQ2.
Specifically, for the three-AI condition, distributions were classified into unanimous agreement (consensus; CON) and two agreeing AIs (divergence; DIV). For the five-AI condition, distributions were classified into unanimous agreement (CON), four agreeing AIs (DIV\_4), and three agreeing AIs (DIV\_3). Because the tasks involved binary classification, these categories exhaustively covered all possible distributions.

\subsection{Results of RQ1: AI panel size}

This section reports the quantitative results for RQ1. AI\_1, AI\_3, and AI\_5 indicate panel sizes of 1, 3, and 5, respectively.

\subsubsection{Decision accuracy}

Figure~\ref{accuracy_AInum} presents decision accuracy by panel size for each of the three tasks. Note that participants’ initial accuracy before viewing any AI predictions was 0.64 on average ($SD = .07$) across all tasks. This suggests that the tasks in our study were understandable yet appropriately challenging. Detailed results are as follows:

\begin{itemize}
    \item \textbf{Income}: AI\_3 ($M \approx .737$) achieved significantly higher accuracy than the single-AI condition ($M \approx .706, p = .012$). AI\_5 did not significantly differ from AI\_1.
    \item \textbf{Recidivism}: A one-way ANOVA found no significant main effect of panel size, $F(2, 81) = 2.33, p = .104$.
    \item \textbf{Dating}: AI\_3 ($Mdn \approx .68$) achieved significantly higher accuracy than AI\_1 ($Mdn \approx .64, p = .002$). AI\_5 was numerically higher than AI\_1 ($p = .064$).
\end{itemize}

Across the Income and Dating tasks, three AIs improved accuracy over a single AI, while five offered no further benefit. This suggests that increasing panel size does not necessarily lead to more accurate decisions.

\subsubsection{Reliance}

Table~\ref{reliance_AInum} summarizes the results for AI reliance indicators across tasks.
Overall, increasing the number of AIs did not yield significant differences for most reliance measures. The only exception was in the Recidivism task, where Agreement Fraction increased. However, since this metric simply captures whether the final answer matches the AI majority~\cite{know_ab_know} and no rise appeared in other indicators, we interpret this result as not indicating increased reliance on AI. As a result, expanding panel size did not consistently increase participants' reliance on AI.

\begin{figure}[ht]
  \centering
  \includegraphics[width=1\linewidth]{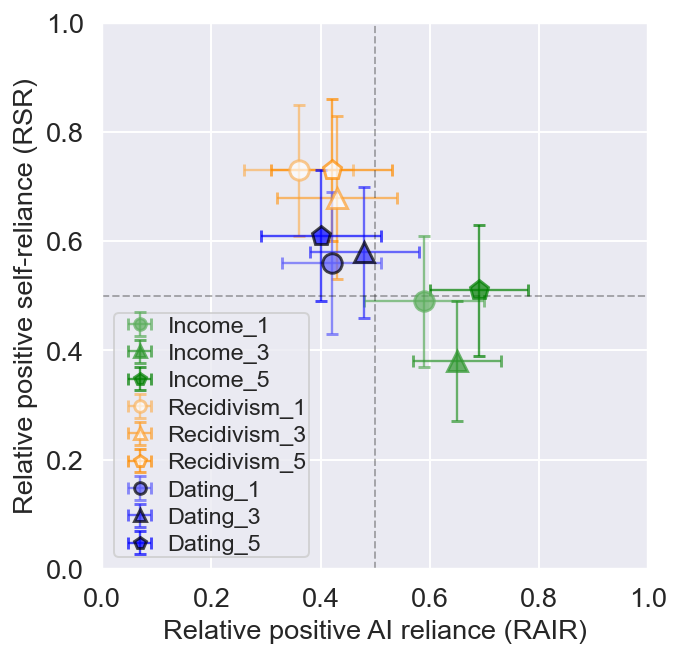}
  \caption{RAIR and RSR across tasks and AI panel sizes. Points show condition means with 95\% confidence intervals. Quadrants indicate reliance patterns: appropriate reliance (top-right), underreliance (top-left), overreliance (bottom-right), and low overall reliance (bottom-left).}
  \Description{Scatter plot of AI reliance (RAIR) and self-reliance (RSR). Each point shows mean values by task and AI panel size with 95\% confidence intervals. Green = Income, orange = Recidivism, blue = Dating; marker shapes indicate 1, 3, or 5 AIs. Dashed lines divide the space into reliance quadrants. Income conditions cluster to the right (greater AI reliance), while Recidivism and Dating cluster higher (greater self-reliance).}
  \label{rair_rsr_AInum}
\end{figure}

Figure\mbox{~\ref{rair_rsr_AInum}} illustrates the relationship between RAIR and RSR across tasks and AI panel sizes. Overall, increasing the number of AI advisors did not systematically change reliance patterns. However, reliance appeared to differ across tasks: it was stronger in Income, while Recidivism showed greater self-reliance due to its high-stakes nature, and Dating showed a similar pattern, potentially reflecting algorithm aversion in intuitive or affective contexts.

\begin{table*}[ht]
  \centering
  \caption{Task results for AI reliance indicators across panel size conditions. Values are reported as ($M \pm SD$).}
  \label{reliance_AInum}
  \begin{tabular}{c|l|c|ccc|c}
    \toprule
    \textbf{Task} & \textbf{Dependent variable} & $p$ & \textbf{AI\_1} & \textbf{AI\_3} & \textbf{AI\_5} & \textbf{Post-hoc results} \\
    \midrule
           & Agreement Fraction & -- & 0.86 $\pm$ 0.10 & 0.89 $\pm$ 0.07 & 0.88 $\pm$ 0.07 & -- \\
           & Switch Fraction    & -- & 0.54 $\pm$ 0.30 & 0.64 $\pm$ 0.22 & 0.60 $\pm$ 0.24 & -- \\
    \textbf{Income} & Accuracy-wid & -- & 0.54 $\pm$ 0.08 & 0.55 $\pm$ 0.12 & 0.60 $\pm$ 0.15 & -- \\
           & RAIR               & -- & 0.59 $\pm$ 0.31 & 0.65 $\pm$ 0.21 & 0.69 $\pm$ 0.24 & -- \\
           & RSR                & -- & 0.49 $\pm$ 0.32 & 0.38 $\pm$ 0.30 & 0.51 $\pm$ 0.30 & -- \\
    \midrule
     & Agreement Fraction & $< .05$ & 0.80 $\pm$ 0.09 & 0.83 $\pm$ 0.12 & 0.85 $\pm$ 0.07 & AI\_1 $<$ AI\_5 \\
                        & Switch Fraction    & --       & 0.33 $\pm$ 0.25 & 0.40 $\pm$ 0.28 & 0.40 $\pm$ 0.26 & -- \\
    \textbf{Recidivism} & Accuracy-wid       & --       & 0.49 $\pm$ 0.12 & 0.49 $\pm$ 0.19 & 0.49 $\pm$ 0.20 & -- \\
                        & RAIR               & --       & 0.36 $\pm$ 0.25 & 0.43 $\pm$ 0.29 & 0.42 $\pm$ 0.27 & -- \\
                        & RSR                & --       & 0.73 $\pm$ 0.30 & 0.68 $\pm$ 0.39 & 0.73 $\pm$ 0.33 & -- \\
    \midrule
    & Agreement Fraction & -- & 0.82 $\pm$ 0.07 & 0.81 $\pm$ 0.08 & 0.79 $\pm$ 0.12 & -- \\
                    & Switch Fraction    & -- & 0.43 $\pm$ 0.22 & 0.45 $\pm$ 0.25 & 0.40 $\pm$ 0.29 & -- \\
    \textbf{Dating} & Accuracy-wid       & -- & 0.46 $\pm$ 0.13 & 0.51 $\pm$ 0.16 & 0.47 $\pm$ 0.16 & -- \\
                    & RAIR               & -- & 0.42 $\pm$ 0.23 & 0.48 $\pm$ 0.28 & 0.40 $\pm$ 0.30 & -- \\
                    & RSR                & -- & 0.56 $\pm$ 0.30 & 0.58 $\pm$ 0.33 & 0.61 $\pm$ 0.34 & -- \\
    \bottomrule
  \end{tabular}
\end{table*}

\subsubsection{Confidence}

Table\mbox{~\ref{confidence_AInum}} shows average confidence ratings before and after AI advice (Initial vs. Final) across tasks and panel sizes. Confidence rose significantly after advice in most conditions. This result indicates that confidence gains came from AI advice itself rather than panel size.

\subsubsection{Subjective measures}

Table~\ref{survey_AInum} summarizes the results of the post-task questionnaire across tasks and AI panel sizes.

Within-task comparisons revealed only one significant effect: in the Income task, ratings on Q4 (autonomy) differed across conditions. Participants in the three-AI condition reported lower autonomy than those in the other conditions ($p < .05$). Open-ended responses reflected this pattern: in the single-AI condition, participants stressed autonomy (e.g., ``I referred to the AI but decided myself''), whereas in the three-AI condition they described stronger AI influence (e.g., ``If all three AIs disagreed, I changed my answer''). In the five-AI condition, participants again emphasized agency (e.g., ``When I was confident, I didn’t change my answer regardless of the AIs''). 
These comments suggest that increasing the number of AIs does not necessarily strengthen psychological reliance on AI; the relationship follows a non-linear shift from autonomy to reliance and back to autonomy.

In contrast, patterns of reliance varied across task conditions. In Income, higher scores in Reliance/Conformity were consistent with quantitative results (Figure~\ref{rair_rsr_AInum}). In Recidivism and Dating, participants stressed that the final decision was their own, using AI only as a reference. Open-ended responses reflected ethical concerns in Recidivism (e.g., Q4: ``I couldn’t let the AI take responsibility'') and algorithm aversion in Dating (e.g., Q6: ``The AI judged mechanically, so it lacked ethics or intelligence'').

In the AI usefulness items (Q7--Q10), no significant differences were observed across panel sizes. Open-ended responses suggest this reflects a balance of benefits and drawbacks: more AIs provided diverse perspectives and reassurance when opinions converged, but disagreement sometimes created confusion. For example, (Q8; Income; AI\_3): ``AIs strengthened my opinion and highlighted new aspects.'' Another noted (Q9; Dating; AI\_3): ``Sometimes the three AIs disagreed, and it confused me.''

Based on prior work\mbox{~\cite{robo_conf_4}}, we estimated participants’ perceived conformity pressure to AI panels by computing Pearson correlations between subjective measures (Q3, Q6) and reliance (Switch Fraction). As shown in Table\mbox{~\ref{correlation_AInum}}, the relationship between psychological influence and decision-making behavior was observed only under multi-AI conditions. Specifically, in Recidivism AI\_5, Switch Fraction was positively correlated with Q3, implying that perceived AI pressure was associated with greater agreement with the AI majority, consistent with normative conformity. In Income AI\_3 and Recidivism AI\_3, perceived AI intelligence was positively correlated with reliance, indicating informational conformity.

Open-ended responses also implied informational conformity, for example, (Q2; Recidivism; AI\_5): ``When the AI repeatedly presented views different from mine, the majority seemed more correct, so I followed it.'' Normative conformity is also reflected in several responses, such as (Q2; Dating; AI\_3): ``When the AI panels formed a majority, I felt I had to go along,'' and (Q3; Recidivism; AI\_5): ``Repeated similar opinions felt pressuring.'' In contrast, some participants described a statistical aggregation strategy, for instance, (Q2; Income; AI\_5): ``When thinking in terms of majority voting, I decided to change my opinion.''

\begin{table*}[ht]
  \centering
  \caption{Initial and Final confidence ($M \pm SD$) across AI panel size conditions for each task (*: $p < .05$, **: $p < .01$).}
  \label{confidence_AInum}
  \begin{tabular}{c|cc|cc|cc}
    \toprule
    \multirow{2}{*}{\textbf{Group}}  & \multicolumn{2}{c}{\textbf{Income}} & \multicolumn{2}{c}{\textbf{Recidivism}} & \multicolumn{2}{c}{\textbf{Dating}} \\
    \cmidrule(lr){2-3} \cmidrule(lr){4-5} \cmidrule(lr){6-7}
     & Initial & Final & Initial & Final & Initial & Final \\
    \midrule
    AI\_1 & 3.81 $\pm$ 0.86 & 3.95 $\pm$ 0.73 & 3.52 $\pm$ 0.78 & \textbf{3.66 $\pm$ 0.74$^{**}$} & 3.50 $\pm$ 0.56 & \textbf{3.81 $\pm$ 0.56$^{**}$} \\
    AI\_3 & 3.74 $\pm$ 0.75 & \textbf{3.99 $\pm$ 0.53$^{**}$} & 3.68 $\pm$ 0.72 & \textbf{3.84 $\pm$ 0.60$^{*}$} & 3.74 $\pm$ 0.49 & \textbf{3.87 $\pm$ 0.52$^{*}$} \\
    AI\_5 & 3.90 $\pm$ 0.64 & \textbf{4.10 $\pm$ 0.54$^{*}$} & 3.59 $\pm$ 0.57 & 3.64 $\pm$ 0.61 & 3.58 $\pm$ 0.80 & \textbf{3.79 $\pm$ 0.61$^{**}$} \\
    \bottomrule
  \end{tabular}
\end{table*}

\begin{table*}[ht]
  \centering
  \caption{Survey results ($Average$) across AI panel sizes for each task (*: $p < .05$).}
  \label{survey_AInum}
  \begin{tabular}{l|l|ccc|ccc|ccc}
    \toprule
    \multirow{2}{*}{\textbf{Category}}  & \multirow{2}{*}{\textbf{No.}} &
    \multicolumn{3}{c}{\textbf{Income}} &
    \multicolumn{3}{c}{\textbf{Recidivism}} &
    \multicolumn{3}{c}{\textbf{Dating}} \\
    \cmidrule(lr){3-5} \cmidrule(lr){6-8} \cmidrule(lr){9-11}
     & & AI\_1 & AI\_3 & AI\_5 & AI\_1 & AI\_3 & AI\_5 & AI\_1 & AI\_3 & AI\_5 \\
    \midrule
    \multirow{6}{*}{\textbf{Reliance \& Conformity}}
     & Q1 & 3.53 & 3.62 & 3.54 & 2.82 & 3.10 & 2.96 & 3.04 & 2.90 & 2.75 \\
     & Q2 & 3.60 & 4.17 & 4.04 & 2.82 & 3.45 & 3.26 & 3.16 & 3.57 & 3.06 \\
     & Q3 & 1.20 & 1.59 & 1.43 & 1.36 & 1.10 & 1.15 & 1.24 & 1.30 & 1.19 \\
     & Q4 & 3.87 & \textbf{3.14$^{*}$} & 4.21 & 4.25 & 4.24 & 4.37 & 4.52 & 4.20 & 4.78 \\
     & Q5 & 1.57 & 1.72 & 1.21 & 1.07 & 0.97 & 1.48 & 1.08 & 0.97 & 1.12 \\
     & Q6 & 3.50 & 4.03 & 3.61 & 3.36 & 3.76 & 3.44 & 2.68 & 2.77 & 2.78 \\
    \midrule
    \multirow{4}{*}{\textbf{AI Usefulness}}
     & Q7  & 2.40 & 2.41 & 2.43 & 1.82 & 2.21 & 2.11 & 2.12 & 2.57 & 2.09 \\
     & Q8  & 4.13 & 4.62 & 4.64 & 4.14 & 4.24 & 3.93 & 4.00 & 3.40 & 3.69 \\
     & Q9  & 4.00 & 4.41 & 4.14 & 3.50 & 3.83 & 3.56 & 3.36 & 3.00 & 3.19 \\
     & Q10 & 4.53 & 4.72 & 4.64 & 3.79 & 4.10 & 3.63 & 3.76 & 3.30 & 4.09 \\
    \bottomrule
  \end{tabular}
\end{table*}

\begin{table*}[ht]
  \centering
  \caption{Pearson correlations ($R$) between subjective conformity measures (Q3, Q6) and reliance (Switch Fraction).}
  \label{correlation_AInum}
  \begin{tabular}{l|c|ccc|ccc|ccc}
    \toprule
    \multirow{2}{*}{\textbf{Question}} & \multirow{2}{*}{\textbf{Statistic}} &
    \multicolumn{3}{c}{\textbf{Income}} &
    \multicolumn{3}{c}{\textbf{Recidivism}} &
    \multicolumn{3}{c}{\textbf{Dating}} \\
    \cmidrule(lr){3-5} \cmidrule(lr){6-8} \cmidrule(lr){9-11}
     & & AI\_1 & AI\_3 & AI\_5 & AI\_1 & AI\_3 & AI\_5 & AI\_1 & AI\_3 & AI\_5 \\
    \midrule
    \textbf{Q3: Pressure}
     & $R$ & 0.107 & 0.076 & 0.250 & -0.020 & 0.178 & 0.422 & 0.038 & 0.116 & 0.093 \\
    \textbf{Normative conformity} & $p$ & -- & -- & -- & -- & -- & $< .05$ & -- & -- & -- \\
    \midrule
    \textbf{Q6: Intelligence}
     & $R$ & 0.238 & -0.079 & 0.478 & 0.261 & 0.425 & 0.023 & -0.049 & 0.296 & 0.096 \\
     \textbf{Informational conformity}& $p$ & -- & -- & $< .01$ & -- & $< .05$ & -- & -- & -- & -- \\
    \bottomrule
  \end{tabular}
\end{table*}

\begin{figure*}[ht]
  \centering
  \begin{subfigure}[b]{0.33\linewidth}
    \centering
    \includegraphics[width=\linewidth]{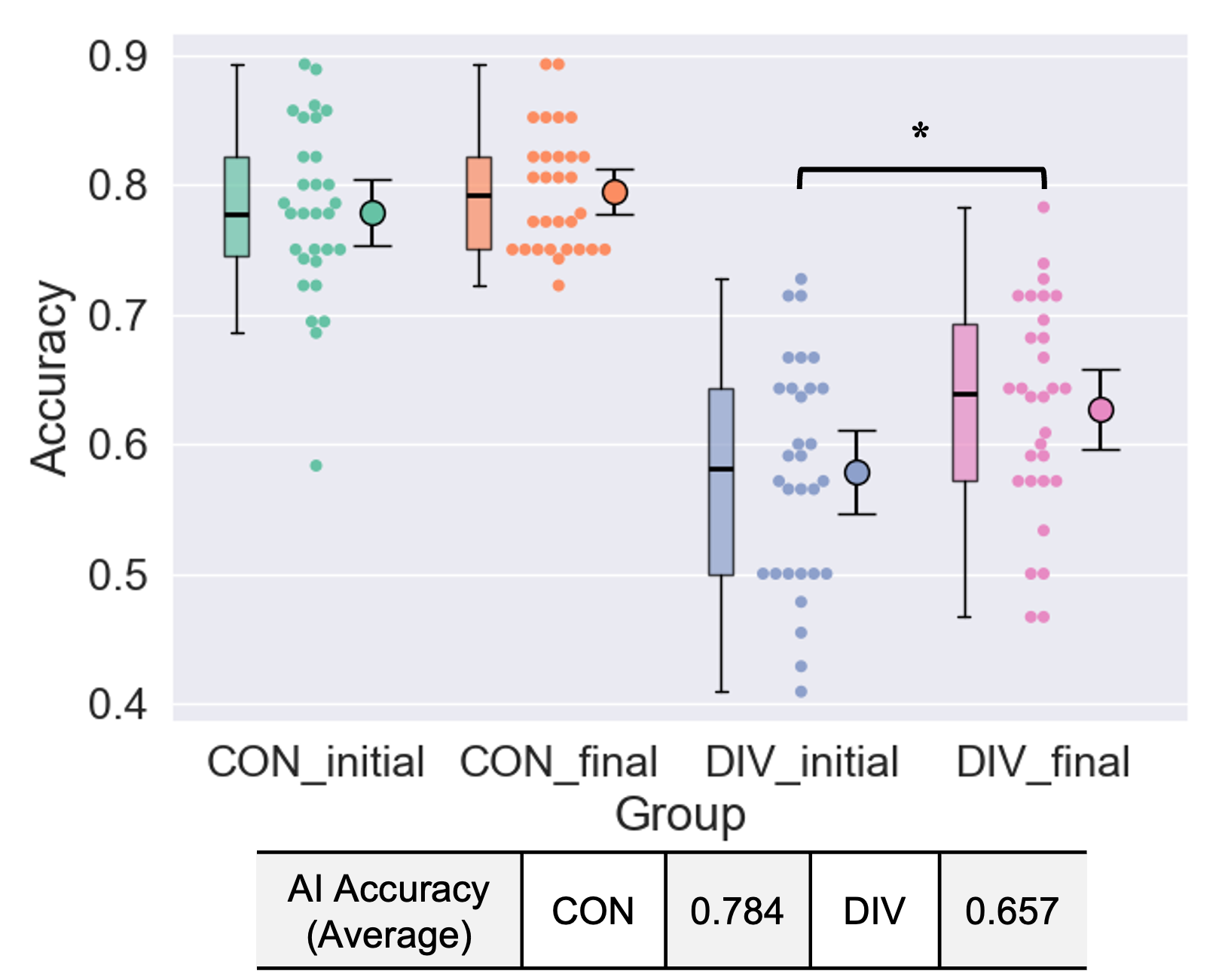}
    \caption{Income}
    \label{accuracy_AIop_a}
  \end{subfigure}
  \begin{subfigure}[b]{0.33\linewidth}
    \centering
    \includegraphics[width=\linewidth]{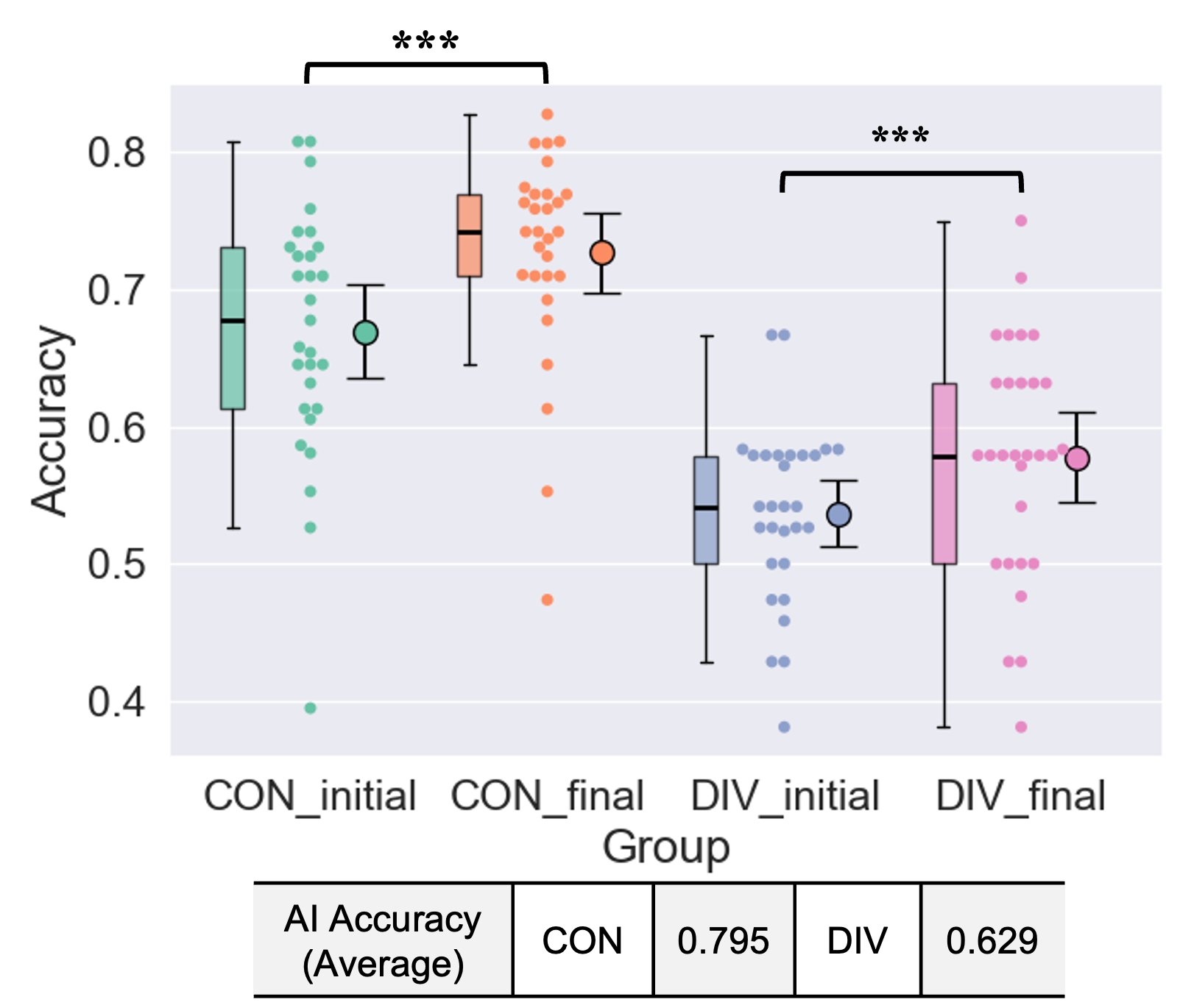}
    \caption{Recidivism}
    \label{accuracy_AIop_b}
  \end{subfigure}
  \begin{subfigure}[b]{0.33\linewidth}
    \centering
    \includegraphics[width=\linewidth]{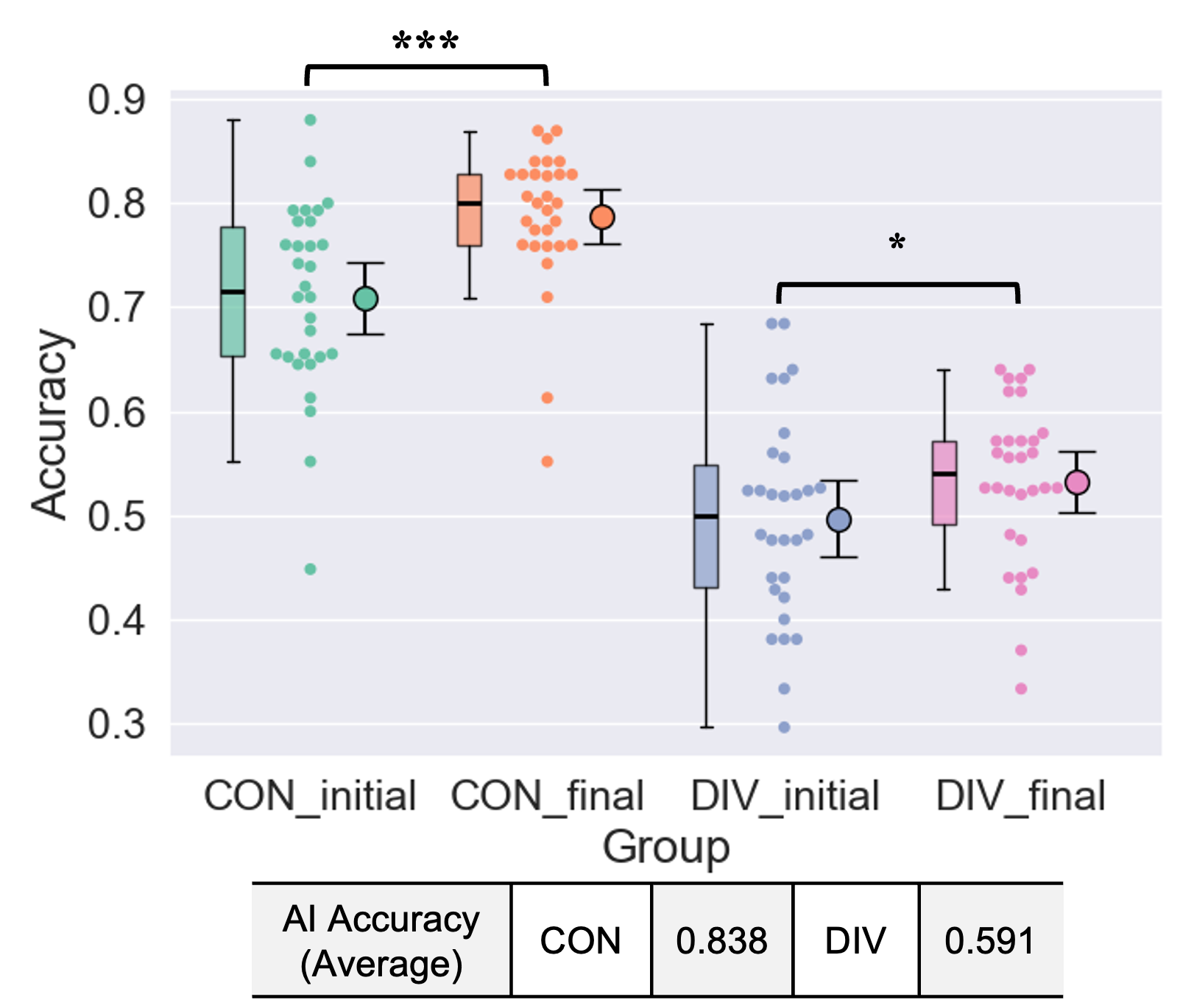}
    \caption{Dating}
    \label{accuracy_AIop_c}
  \end{subfigure}
  \caption{Decision accuracy across consensus (CON) and divergence (DIV) conditions when the AI panel size was three (*: $p < .05$, **: $p < .01$, ***: $p < .001$). Decision accuracy is plotted separately for initial responses before viewing AI predictions (CON\_initial, DIV\_initial) and final responses after viewing AI predictions (CON\_final, DIV\_final). The tables beneath each panel report the average accuracy of the majority AI prediction.}
  \Description{Decision accuracy under consensus and divergent AI advice. Boxplots show accuracy in Income, Recidivism, and Dating tasks when three AIs gave unanimous (CON) or divergent (DIV) predictions. Each panel compares initial answers with final answers after consulting AI. Distributions, individual scores, and mean values with 95\% confidence intervals are plotted. Tables below each panel report majority AI accuracy, showing that consensus trials were generally easier. Accuracy was higher in consensus than divergence, and most conditions showed improvement after AI advice.}
  \label{accuracy_AIop}
\end{figure*}

\begin{figure*}[ht]
  \centering
  \begin{subfigure}[b]{0.33\linewidth}
    \centering
    \includegraphics[width=\linewidth]{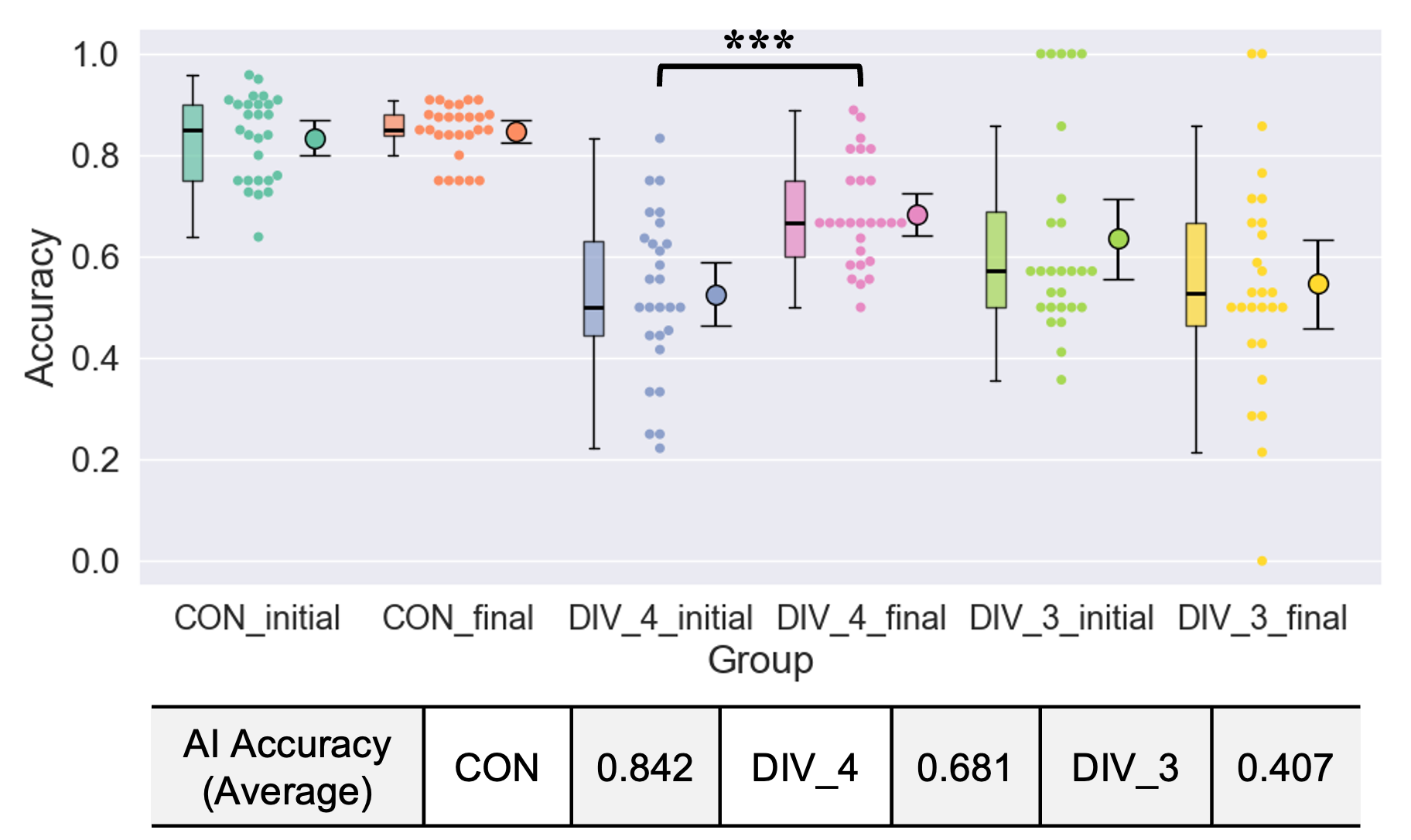}
    \caption{Income}
    \label{accuracy_AIop5_a}
  \end{subfigure}
  \begin{subfigure}[b]{0.33\linewidth}
    \centering
    \includegraphics[width=\linewidth]{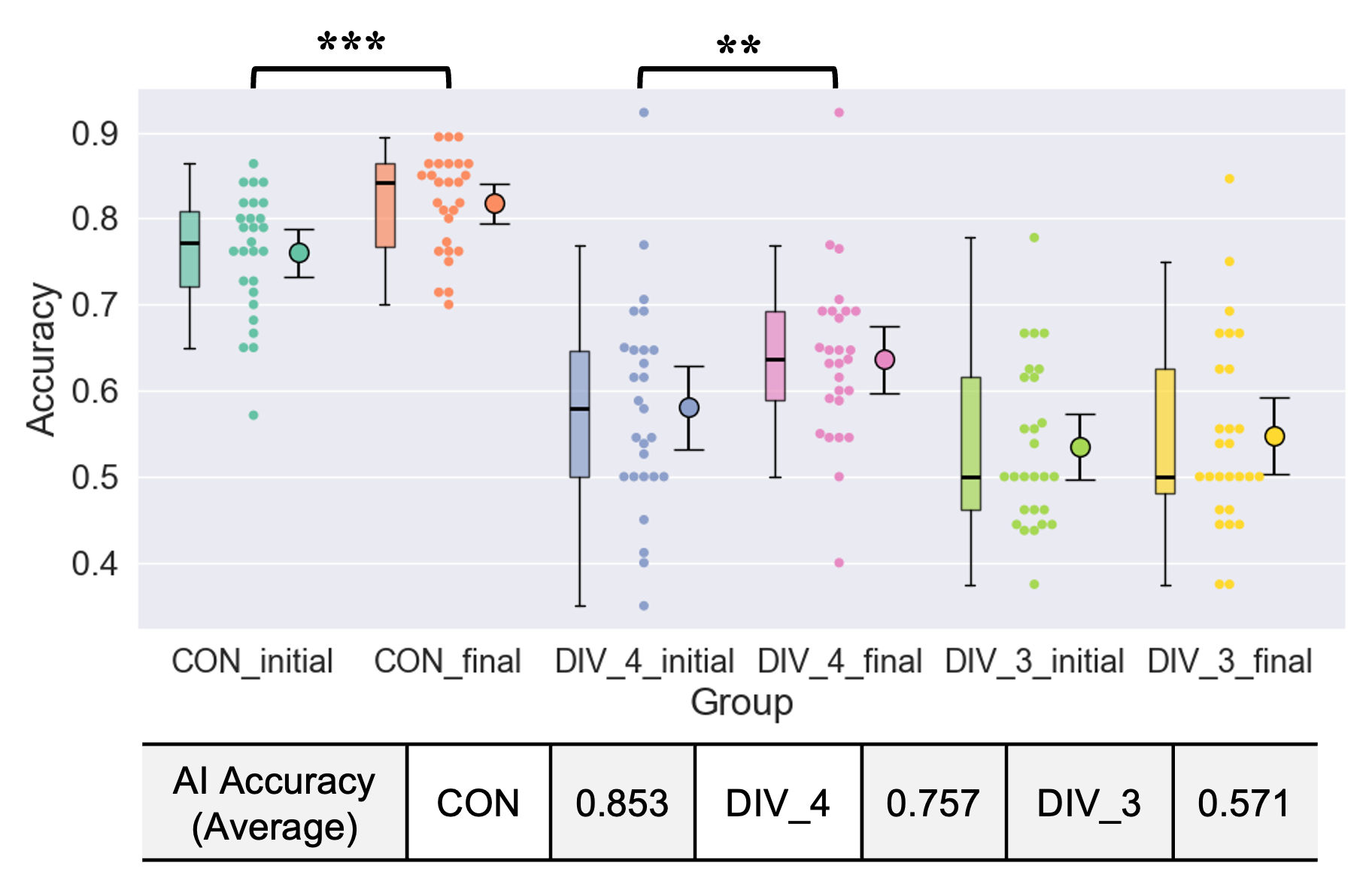}
    \caption{Recidivism}
    \label{accuracy_AIop5_b}
  \end{subfigure}
  \begin{subfigure}[b]{0.33\linewidth}
    \centering
    \includegraphics[width=\linewidth]{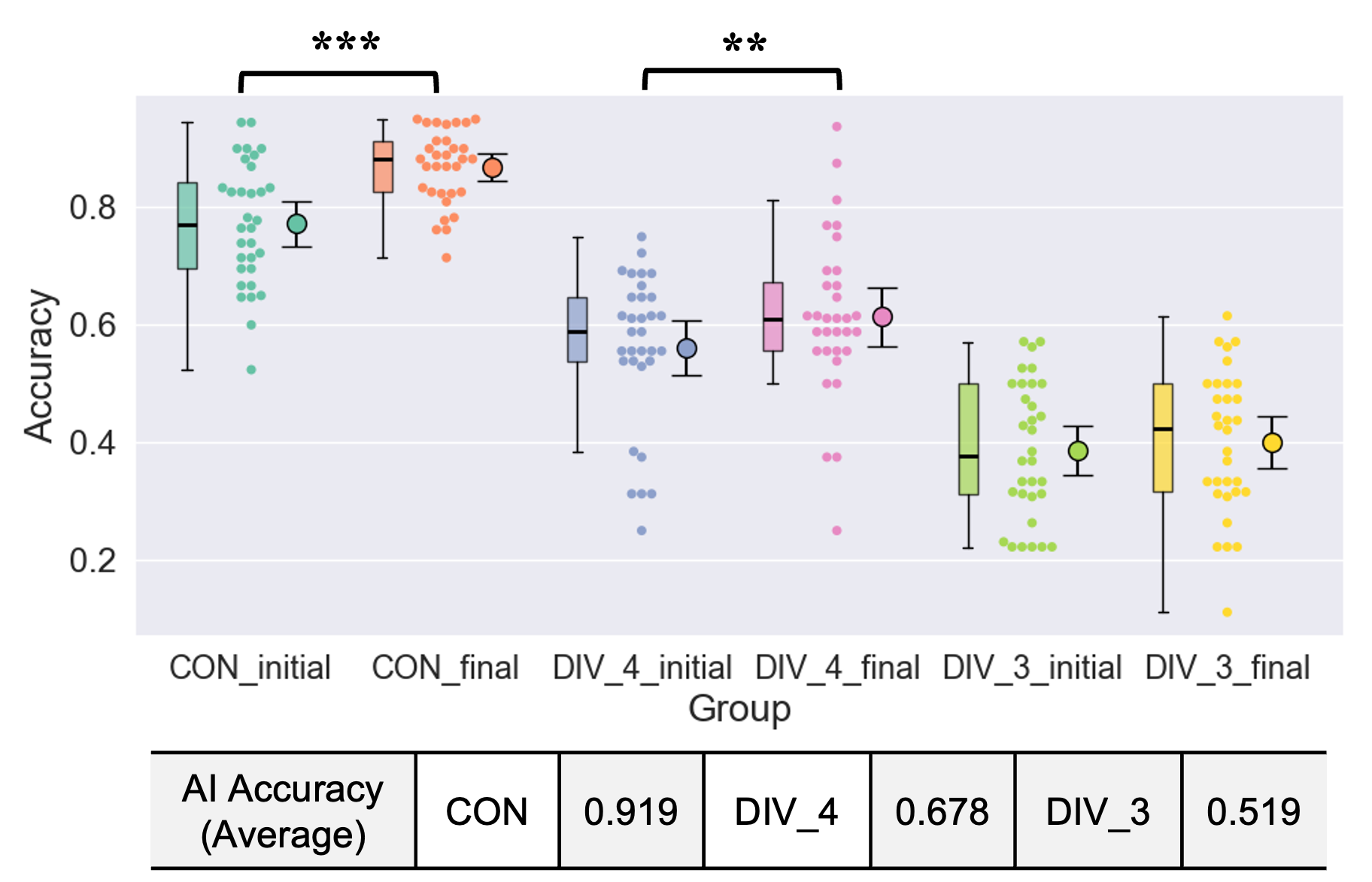}
    \caption{Dating}
    \label{accuracy_AIop5_c}
  \end{subfigure}
  \caption{Accuracy across opinion distributions in the five panels condition (CON, DIV\_4, and DIV\_3). Accuracy is compared before (Initial) and after (Final) viewing AI predictions (*: $p < .05$, **: $p < .01$, ***: $p < .001$).}
  \Description{Decision accuracy with five AIs across consensus and split conditions. Boxplots show initial and final accuracy in Income, Recidivism, and Dating tasks for unanimous (CON), 4--1 (DIV_4), and 3--2 (DIV_3) opinion splits. Distributions, individual scores, and means with 95\% confidence intervals are plotted, with tables below showing majority AI accuracy. Accuracy was highest in consensus trials; DIV_4 showed gains from clear majorities, while DIV_3 showed no improvement due to ambiguity.}
  \label{accuracy_AIop5}
\end{figure*}

\subsection{Results of RQ2: within-panel consensus}

\subsubsection{Decision accuracy}

\paragraph{Three panels}

Figure\mbox{~\ref{accuracy_AIop}} shows decision accuracy across opinion distributions in the three-AI condition. We compared pre- vs. post-advice accuracy within each opinion distribution. In the DIV conditions, post-advice accuracy was significantly higher than pre-advice accuracy across all tasks. In the CON conditions, no improvement was observed for Income, likely because pre-advice accuracy was already high. By contrast, for Recidivism and Dating, accuracy increased from pre- to post-advice ($p$ < .001). Therefore, when three AIs were present, decision accuracy improved after consulting AI advice in most cases, regardless of whether the AI panel reached consensus or showed divergence.

\begin{table*}[ht]
  \centering
  \caption{Comparison of AI reliance measures in the AI\_3 condition ($M \pm SD$). The table summarizes within-task comparison results across CON and DIV trials (*: $p < .05$, **: $p < .01$, ***: $p < .001$).}
  \label{reliance_AIop}
  \begin{tabular}{l|cc|cc|cc}
    \toprule
    \multirow{2}{*}{\textbf{Dependent variable}} &
    \multicolumn{2}{c|}{\textbf{Income}} &
    \multicolumn{2}{c|}{\textbf{Recidivism}} &
    \multicolumn{2}{c}{\textbf{Dating}} \\
    \cmidrule(lr){2-3} \cmidrule(lr){4-5} \cmidrule(lr){6-7}
     & CON & DIV & CON & DIV & CON & DIV \\
    \midrule
    Agreement Fraction & \textbf{0.97 $\pm$ 0.05$^{***}$} & 0.73 $\pm$ 0.12 & \textbf{0.91 $\pm$ 0.12$^{***}$} & 0.70 $\pm$ 0.19 & \textbf{0.93 $\pm$ 0.08$^{***}$} & 0.67 $\pm$ 0.12 \\
    Switch Fraction    & \textbf{0.84 $\pm$ 0.27$^{***}$} & 0.44 $\pm$ 0.28 & \textbf{0.56 $\pm$ 0.34$^{**}$}  & 0.28 $\pm$ 0.32 & \textbf{0.63 $\pm$ 0.33$^{***}$} & 0.35 $\pm$ 0.24 \\
    Accuracy-wid       & \textbf{0.57 $\pm$ 0.21$^{*}$} & 0.51 $\pm$ 0.15 & 0.53 $\pm$ 0.31 & 0.47 $\pm$ 0.19 & 0.57 $\pm$ 0.28 & 0.48 $\pm$ 0.16 \\
    RAIR               & \textbf{0.90 $\pm$ 0.22$^{***}$} & 0.46 $\pm$ 0.31 & \textbf{0.54 $\pm$ 0.33$^{**}$}  & 0.31 $\pm$ 0.33 & \textbf{0.59 $\pm$ 0.37$^{*}$}   & 0.37 $\pm$ 0.25 \\
    RSR                & \textbf{0.21 $\pm$ 0.36$^{***}$} & 0.60 $\pm$ 0.31 & \textbf{0.42 $\pm$ 0.51$^{*}$} & 0.75 $\pm$ 0.39 & \textbf{0.32 $\pm$ 0.42$^{**}$}  & 0.65 $\pm$ 0.33 \\
    \bottomrule
  \end{tabular}
\end{table*}

\begin{table*}[ht]
  \centering
  \caption{Comparison of AI reliance indices under CON, DIV\_4, and DIV\_3 conditions for three tasks. 
  Values are reported as $M \pm SD$. Post-hoc comparisons indicate significant pairwise differences (Bonferroni corrected).}
  \label{reliance_AIop5}
  \begin{tabular}{l|l|c|ccc|c}
    \toprule
    \textbf{Task} & \textbf{Dependent variable} & $p$ & \textbf{CON} & \textbf{DIV\_4} & \textbf{DIV\_3} & \textbf{Post-hoc results} \\
    \midrule
    \multirow{5}{*}{\textbf{Income}} 
      & Agreement Fraction & $< .001$ & 0.99 $\pm$ 0.03 & 0.90 $\pm$ 0.13 & 0.53 $\pm$ 0.23 & CON, DIV\_4 $>$ DIV\_3 \\
      & Switch Fraction    & $< .001$ & 0.88 $\pm$ 0.28 & 0.74 $\pm$ 0.30 & 0.30 $\pm$ 0.32 & CON, DIV\_4 $>$ DIV\_3 \\
      & Accuracy-wid       & -- & 0.56 $\pm$ 0.35 & 0.75 $\pm$ 0.20 & 0.56 $\pm$ 0.32 & -- \\
      & RAIR               & $< .001$ & 0.99 $\pm$ 0.03 & 0.80 $\pm$ 0.28 & 0.31 $\pm$ 0.33 & CON, DIV\_4 $>$ DIV\_3 \\
      & RSR                & $< .001$ & 0.20 $\pm$ 0.39 & 0.51 $\pm$ 0.46 & 0.68 $\pm$ 0.35 & CON $<$ DIV\_3 \\
    \midrule
    \multirow{5}{*}{\textbf{Recidivism}}
      & Agreement Fraction & $< .001$ & 0.94 $\pm$ 0.06 & 0.85 $\pm$ 0.09 & 0.67 $\pm$ 0.20 & CON $>$ DIV\_4 $>$ DIV\_3 \\
      & Switch Fraction    & $< .001$ & 0.67 $\pm$ 0.35 & 0.38 $\pm$ 0.33 & 0.21 $\pm$ 0.28 & CON $>$ DIV\_4 $>$ DIV\_3 \\
      & Accuracy-wid       & $< .01$  & 0.67 $\pm$ 0.34 & 0.37 $\pm$ 0.29 & 0.45 $\pm$ 0.24 & CON $>$ DIV\_4, DIV\_3 \\
      & RAIR               & $< .001$ & 0.68 $\pm$ 0.36 & 0.35 $\pm$ 0.34 & 0.22 $\pm$ 0.31 & CON $>$ DIV\_4, DIV\_3 \\
      & RSR                & --       & 0.58 $\pm$ 0.50 & 0.50 $\pm$ 0.53 & 0.84 $\pm$ 0.25 & -- \\
    \midrule
    \multirow{5}{*}{\textbf{Dating}}
      & Agreement Fraction & $< .001$ & 0.94 $\pm$ 0.06 & 0.79 $\pm$ 0.16 & 0.61 $\pm$ 0.19 & CON $>$ DIV\_4 $>$ DIV\_3 \\
      & Switch Fraction    & $< .001$ & 0.65 $\pm$ 0.37 & 0.42 $\pm$ 0.37 & 0.28 $\pm$ 0.30 & CON $>$ DIV\_4 $>$ DIV\_3 \\
      & Accuracy-wid       & $< .01$  & 0.62 $\pm$ 0.34 & 0.49 $\pm$ 0.24 & 0.38 $\pm$ 0.22 & CON $>$ DIV\_3 \\
      & RAIR               & $< .001$ & 0.63 $\pm$ 0.38 & 0.43 $\pm$ 0.39 & 0.24 $\pm$ 0.33 & CON, DIV\_4 $>$ DIV\_3 \\
      & RSR                & --       & 0.44 $\pm$ 0.53 & 0.62 $\pm$ 0.45 & 0.67 $\pm$ 0.39 & -- \\
    \bottomrule
  \end{tabular}%
\end{table*}

\begin{figure}[ht]
  \centering
  \includegraphics[width=1\linewidth]{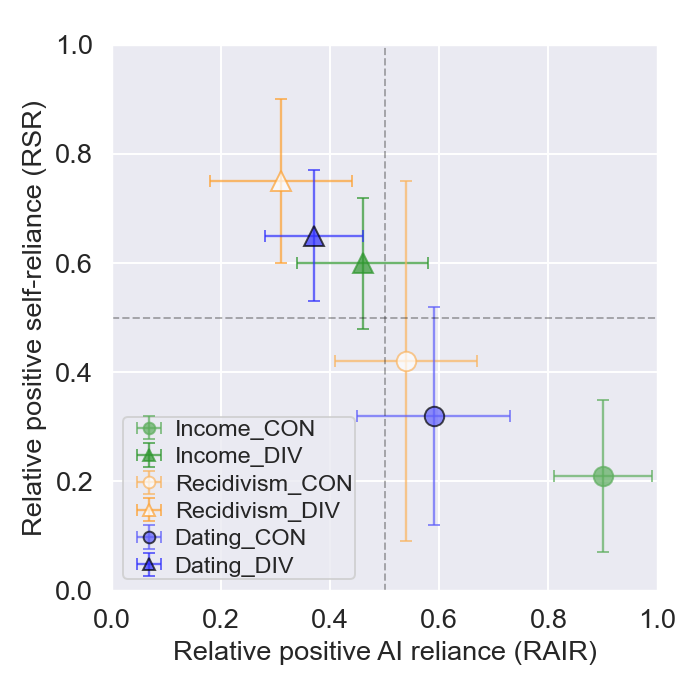}
  \caption{RAIR and RSR for each task under the three panels condition, separated into CON and DIV cases.}
  \Description{Scatter plot of RAIR (x-axis) and RSR (y-axis) with 95\% confidence intervals. Colors indicate tasks (Income = green, Recidivism = orange, Dating = blue); circles mark consensus and triangles mark divergence. Consensus points lie further right, especially for Income, showing stronger AI reliance, while divergence points in Recidivism and Dating are higher, indicating stronger self-reliance. Dashed lines divide quadrants of over, under, and appropriate reliance.
  }
  \label{rair_rsr_AIop}
\end{figure}

\begin{figure}[htbp]
  \centering
  \includegraphics[width=1\linewidth]{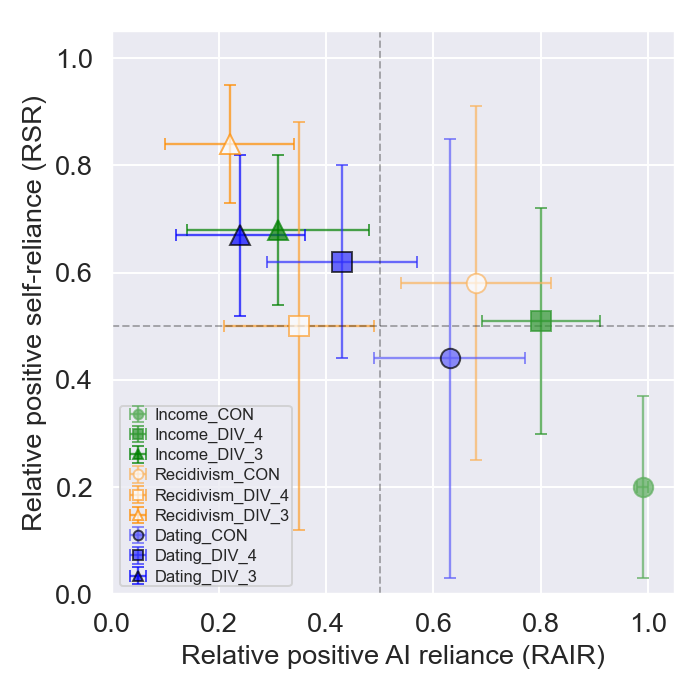}
  \caption{RAIR and RSR under the five panels condition (CON, DIV\_4, DIV\_3).}
  \Description{Scatter plot of RAIR (x-axis) and RSR (y-axis) with 95\% confidence intervals. Colors show tasks (green = Income, orange = Recidivism, blue = Dating); shapes show opinion distributions (circle = CON, square = DIV_4, triangle = DIV_3). Consensus clusters lower right, reflecting stronger AI reliance, while DIV_3 appears more balanced or self-reliant. The figure shows that reliance patterns vary by task and opinion distribution even with five AIs.}
  \label{rair_rsr_AIop5}
\end{figure}

\paragraph{Five panels}

Figure~\ref{accuracy_AIop5} presents decision accuracy across opinion distributions in the five-AI condition. When all AIs agreed, the AI predictions were highly accurate, and participants also achieved high accuracy by agreeing with AIs. However, differences emerged in cases of partial agreement compared to the three-AI condition. In DIV\_4, post-advice accuracy was significantly higher than pre-advice accuracy across all tasks ($ps$ < .01). By contrast, in DIV\_3, no significant pre-post accuracy differences were found in any task. This suggests that a near-even split among AI opinions created uncertainty and undermined effective decision-making.

\subsubsection{Reliance}

\paragraph{Three panels}

Table~\ref{reliance_AIop} reports the comparison of AI reliance between CON and DIV trials in the three-AI condition. Agreement and Switch Fractions were consistently higher in CON than in DIV, indicating that participants often revised their answers to follow consensus. This effect was strongest in Income, whereas in Recidivism and Dating they more often retained their own judgments, likely due to the high-stakes or affective nature of these tasks~\cite{alg_aversion}.
For Accuracy-wid, no significant differences were observed between CON and DIV conditions. However, as shown in Figure~\ref{rair_rsr_AIop}, RAIR increased under CON in all tasks, indicating stronger reliance on AI when consensus was present. Conversely, RSR also increased under DIV in all tasks, reflecting cases where participants maintained their own correct initial responses in the face of disagreement.

\paragraph{Five panels}

Table~\ref{reliance_AIop5} reports the reliance indicators by opinion distribution in the five-AI condition, and Figure~\ref{rair_rsr_AIop5} illustrates the relationship between RAIR and RSR.
As shown in Table~\ref{reliance_AIop5}, both Agreement and Switch Fractions were highest in the CON condition across all tasks, indicating strong conformity pressure and overreliance on the AI's consensus. By contrast, in the DIV\_4 condition, Agreement and Switch Fractions were significantly lower than in CON for the Recidivism and Dating tasks. These findings suggest that the presence of even a single dissenting opinion partially alleviated conformity pressure from AI panels.

For Accuracy-wid and RAIR, scores were significantly higher in CON for the Recidivism and Dating tasks, indicating that reliance on AI consensus contributed positively to accuracy in these contexts. However, in DIV\_3, RAIR was significantly lower, while RSR did not show a significant difference. These results suggests that within-panel disagreement can hinder reliance on the AI majority but does not necessarily lead to more accurate decisions.

\begin{table*}[ht]
  \centering
  \caption{Initial and Final confidence ($M \pm SD$) under CON and DIV conditions across three tasks ($^{*}p<.05$, $^{***}p<.001$).}
  \label{confidence_AIop}
  \begin{tabular}{l|cc|cc|cc}
    \toprule
    \multirow{2}{*}{\textbf{Group}} &     \multicolumn{2}{c|}{\textbf{Income}} &
    \multicolumn{2}{c|}{\textbf{Recidivism}} &
    \multicolumn{2}{c}{\textbf{Dating}} \\
    \cmidrule(lr){2-3} \cmidrule(lr){4-5} \cmidrule(lr){6-7}
    & Initial & Final & Initial & Final & Initial & Final \\
    \midrule
    CON & 3.90 $\pm$ 0.71 & \textbf{4.41 $\pm$ 0.46$^{***}$} & 
          3.66 $\pm$ 0.77 & \textbf{4.06 $\pm$ 0.69$^{***}$} &
          3.86 $\pm$ 0.52 & \textbf{4.26 $\pm$ 0.51$^{***}$} \\
    DIV & 3.47 $\pm$ 0.87 & \textbf{3.26 $\pm$ 0.72$^{*}$} &
          3.69 $\pm$ 0.78 & \textbf{3.46 $\pm$ 0.65$^{*}$} &
          3.58 $\pm$ 0.57 & \textbf{3.36 $\pm$ 0.76$^{*}$} \\
    \bottomrule
  \end{tabular}
\end{table*}

\begin{table*}[ht]
  \centering
  \caption{Initial and Final confidence ($M \pm SD$) under CON, DIV\_4, and DIV\_3 for three tasks (*: $p < .05$, **: $p < .01$, ***: $p < .001$).}
  \label{confidence_AIop5}
  \begin{tabular}{l|cc|cc|cc}
    \toprule
    \multirow{2}{*}{\textbf{Group}} &
    \multicolumn{2}{c|}{\textbf{Income}} &
    \multicolumn{2}{c|}{\textbf{Recidivism}} &
    \multicolumn{2}{c}{\textbf{Dating}} \\
    \cmidrule(lr){2-3} \cmidrule(lr){4-5} \cmidrule(lr){6-7}
     & Initial & Final & Initial & Final & Initial & Final \\
    \midrule
    CON & 
    4.16 $\pm$ 0.61 & \textbf{4.59 $\pm$ 0.52$^{***}$} &
    3.57 $\pm$ 0.61 & \textbf{3.91 $\pm$ 0.67$^{***}$} &
    3.81 $\pm$ 0.75 & \textbf{4.30 $\pm$ 0.44$^{***}$} \\
    DIV\_4 & 
    3.65 $\pm$ 0.79 & 3.87 $\pm$ 0.70 &
    3.61 $\pm$ 0.56 & 3.62 $\pm$ 0.64 &
    3.50 $\pm$ 0.85 & 3.65 $\pm$0.67 \\
    DIV\_3 &
    3.79 $\pm$ 0.80 & \textbf{3.17 $\pm$ 1.00 $^{**}$} &
    3.54 $\pm$ 0.81 & \textbf{3.13 $\pm$ 0.95 $^{**}$} &
    3.35 $\pm$ 0.97 & 3.21 $\pm$ 1.05\\
    \bottomrule
  \end{tabular}
\end{table*}

\subsubsection{Confidence}

Table\mbox{~\ref{confidence_AIop}} and Table\mbox{~\ref{confidence_AIop5}} show changes in participants’ confidence from initial to final judgments across different panel sizes. In both the three- and five-AI conditions, confidence increased under CON, reflecting strong reinforcement from unanimous panel consensus. By contrast, disagreement within the panel induced confidence drops under DIV conditions, suggesting that even a single dissenting AI opinion can prompt greater caution in decision-making.

\section{Study 2}

\subsection{Setup}

In Study 2, we manipulated the human-likeness of the AI panels as a between-subjects factor to examine RQ3. The following two conditions were tested:
\begin{itemize}
\item \textbf{Standard AI}: Three-AI panel identical to that used in Study 1; the same participant data were used.
\item \textbf{Human-like AI}: Three-AI panel augmented with facial photos, names, and conversational tone. For this condition, new participants were recruited to ensure no overlap with Study 1.
\end{itemize}

New participants were randomly assigned to one of three tasks and then to one of the AI conditions. The number of participants per task for the human-like AI condition was as follows: Income = 30, Recidivism = 27, and Dating = 31.

An example of the generated agent interface is shown in Figure~\ref{interface2}. We incorporated three commonly used elements for AI anthropomorphism~\cite{chat_medi,anth_same1,multi_norms}: (1) visual human-like appearance, (2) human names and identities, and (3) natural conversational style.  These elements were implemented in two ways:

\begin{figure}[htbp]
  \centering
  \includegraphics[width=\linewidth]{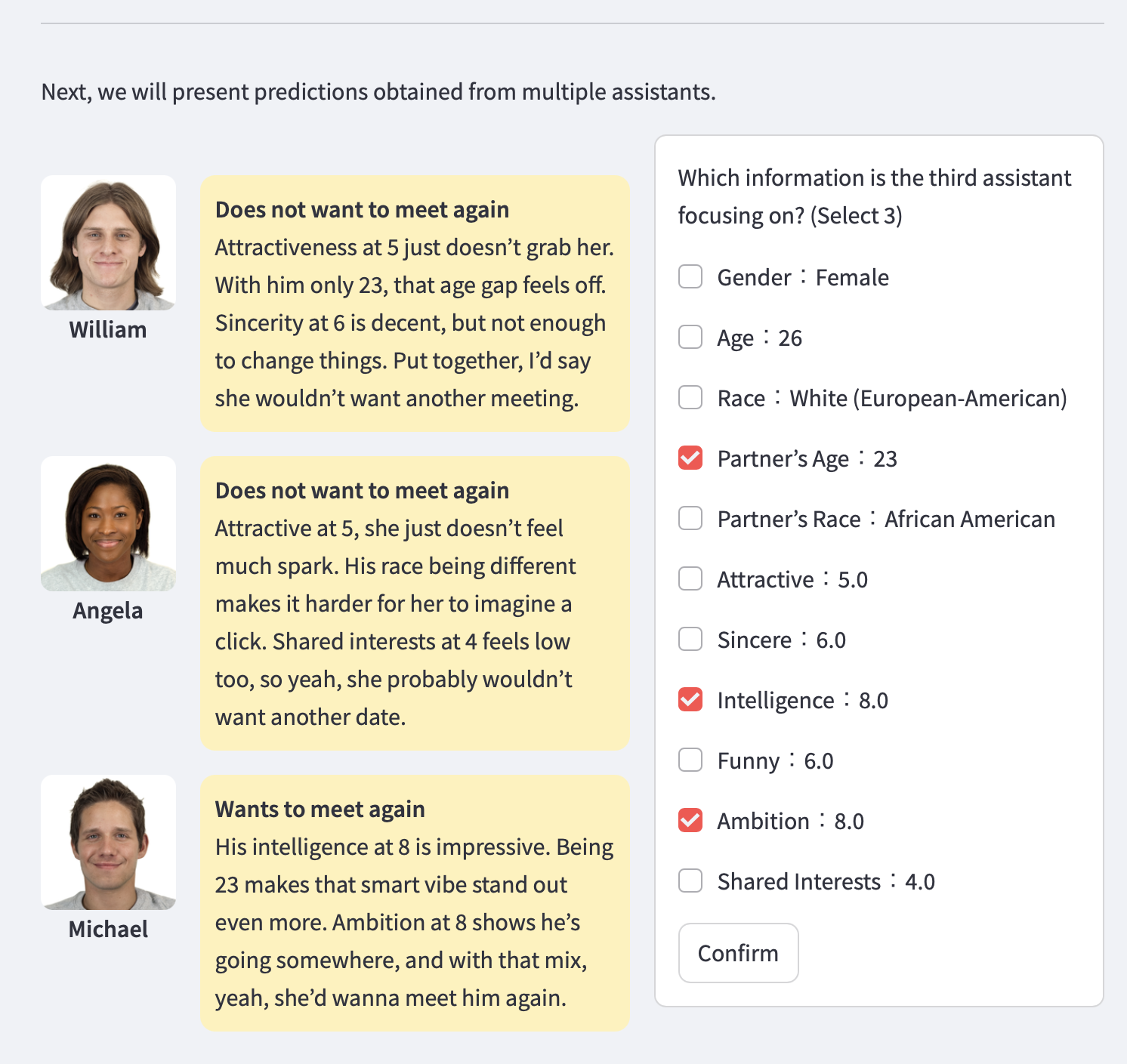}
  \caption{Example of the AI advice and attention check screen in the Dating task under the human-like advisor condition. Three AIs are presented with (1) human-like visuals, (2) personal names/identities, and (3) conversational, natural-language styles. Each agent provides a prediction about whether the participant would want to meet the partner again, along with a rationale. To ensure careful reading, participants are asked to select the features that the third assistant is focusing on from a checklist.}
  \Description{Interface for the Dating task with human-like AI advisors. Three advisors appear with photos, names, binary predictions, and natural-language explanations. Each emphasizes different profile attributes (e.g., age, race, intelligence). On the right, an attention-check panel asks participants to select the attributes mentioned by the third advisor.}
  \label{interface2}
\end{figure}

\begin{figure*}[ht]
  \centering
  \begin{subfigure}[b]{0.33\linewidth}
    \centering
    \includegraphics[width=\linewidth]{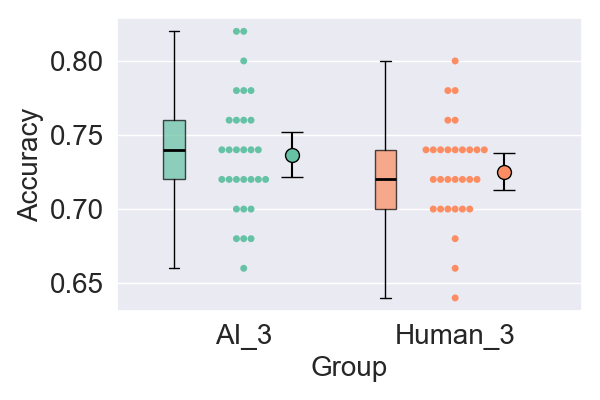}
    \caption{Income}
    \label{accuracy_AIH_a}
  \end{subfigure}
  \begin{subfigure}[b]{0.33\linewidth}
    \centering
    \includegraphics[width=\linewidth]{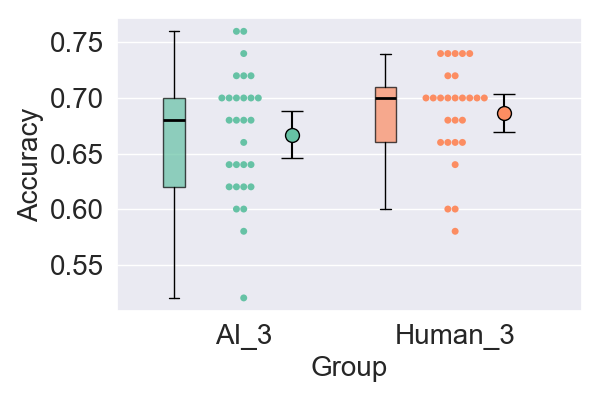}
    \caption{Recidivism}
    \label{accuracy_AIH_b}
  \end{subfigure}
  \begin{subfigure}[b]{0.33\linewidth}
    \centering
    \includegraphics[width=\linewidth]{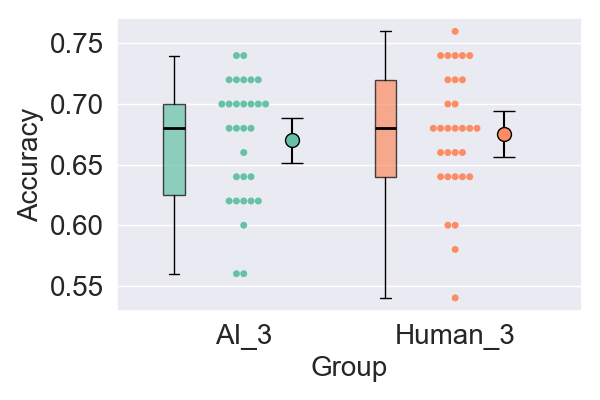}
    \caption{Dating}
    \label{accuracy_AIH_c}
  \end{subfigure}
  \caption{Comparison of decision accuracy between AI advisors (AI\_3) and human-like advisors (Human\_3). Distributions are shown with boxplots, individual data points, and mean values with 95\% confidence intervals. Across all tasks, no significant differences were found between AI\_3 and Human\_3 conditions (all $p$ > .05).}
  \Description{Decision accuracy with AI versus human-like advisors. Boxplots show accuracy in Income, Recidivism, and Dating tasks when advice came from three AIs (AI\_3) or three human-like advisors (Human\_3). Distributions, individual scores, and mean accuracy with 95\% confidence intervals are plotted. Accuracy patterns are similar across conditions in all tasks.}
  \label{accuracy_AIH}
\end{figure*}

\begin{table*}[ht]
  \centering
  \caption{Comparison of AI reliance between AI\_3 and Human\_3 across tasks ($M \pm SD$).}
  \label{reliance_AIH}
  \begin{tabular}{l|cc|cc|cc}
    \toprule
    \multirow{2}{*}{\textbf{Dependent variable}} &
    \multicolumn{2}{c|}{\textbf{Income}} &
    \multicolumn{2}{c|}{\textbf{Recidivism}} &
    \multicolumn{2}{c}{\textbf{Dating}} \\
    \cmidrule(lr){2-3} \cmidrule(lr){4-5} \cmidrule(lr){6-7}
     & AI\_3 & Human\_3 & AI\_3 & Human\_3 & AI\_3 & Human\_3 \\
    \midrule
    Agreement Fraction & 0.89 $\pm$ 0.07 & 0.88 $\pm$ 0.10 & 0.83 $\pm$ 0.12 & 0.83 $\pm$ 0.09 & 0.81 $\pm$ 0.08 & 0.82 $\pm$ 0.10 \\
    Switch Fraction    & 0.64 $\pm$ 0.22 & 0.63 $\pm$ 0.25 & 0.40 $\pm$ 0.28 & 0.41 $\pm$ 0.25 & 0.45 $\pm$ 0.25 & 0.52 $\pm$ 0.26 \\
    Accuracy-wid       & 0.55 $\pm$ 0.12 & 0.57 $\pm$ 0.16 & 0.49 $\pm$ 0.19 & 0.53 $\pm$ 0.16 & 0.51 $\pm$ 0.16 & 0.56 $\pm$ 0.13 \\
    RAIR               & 0.65 $\pm$ 0.21 & 0.64 $\pm$ 0.29 & 0.43 $\pm$ 0.29 & 0.46 $\pm$ 0.25 & 0.48 $\pm$ 0.28 & 0.55 $\pm$ 0.25 \\
    RSR                & 0.38 $\pm$ 0.30 & 0.43 $\pm$ 0.31 & 0.68 $\pm$ 0.39 & 0.74 $\pm$ 0.36 & 0.58 $\pm$ 0.33 & 0.56 $\pm$ 0.34 \\
    \bottomrule
  \end{tabular}
\end{table*}

\begin{itemize}
\item \textbf{Facial photos and names}: We used human face images drawn from the Chicago Face Database~\cite{ma2015chicago} and assigned names randomly generated by GPT-4o. The images included both male and female faces, as well as racially diverse backgrounds (e.g., Asian, Black, White).
\item \textbf{Conversational tone}: In prompts to the AI (Figure~\ref{prompt}), we appended the instruction: \textit{Please role-play as the assigned persona and generate your response in a conversational style.}
\end{itemize}

At the beginning of each task, three AIs were randomly sampled from the Rashomon set and paired with a pre-assigned face, name, and conversational tone. We required that each panel include at least one male and one female agent, thereby isolating the effect of perceived human-likeness rather than gender differences~\cite{gender}.

Furthermore, to verify whether our manipulation successfully induced participants’ perceptions of human-likeness, we conducted a manipulation check on the same platform. The results showed significant differences in multiple anthropomorphism-related items, i.e., perceiving the advisors’ speaking style as more human-like, feeling that they had their own intentions, and sensing emotions or personality. Details are provided in Appendix\mbox{~\ref{check_sec}}.

\subsection{Results of RQ3: human-likeness}

As in Study 1, we report results on accuracy, AI reliance, confidence, and subjective evaluations. Comparisons between the human-like AI panels (Human\_3) and standard-AI panels (AI\_3) were conducted using independent two-sample tests (t-tests or Wilcoxon rank-sum tests, depending on normality).

\subsubsection{Decision accuracy}

Figure~\ref{accuracy_AIH} compares decision accuracy between AI\_3 and Human\_3. Independent two-sample comparisons showed no significant differences in decision accuracy across tasks ($ps$ > .05), indicating no effect of the human-like presentation.

\subsubsection{Reliance}

Table~\ref{reliance_AIH} reports the comparison of reliance indicators between AI\_3 and Human\_3.
Across all tasks, no significant differences were observed between the AI and Human conditions, indicating that the advisors' human-like appearance did not systematically affect reliance measures.

\subsubsection{Confidence}

Table~\ref{confidence_AIH} presents the change in confidence in the Human~\_3 condition. Paired-sample t-tests or Wilcoxon signed-rank tests showed that participants’ confidence consistently increased after consulting the human-like advisors across all tasks, as in the AI~\_3 condition (Table~\ref{confidence_AInum}).

\begin{table}[htbp]
  \centering
  \caption{Initial and Final results of confidence ($M \pm SD$) for Human\_3 across tasks.}
  \label{confidence_AIH}
  \begin{tabular}{l|c|cc}
    \toprule
    \textbf{Task} & $p$ & \textbf{Initial} & \textbf{Final} \\
    \midrule
    Income & $<0.001$ & 3.56 $\pm$ 0.74 & \textbf{3.85 $\pm$ 0.63$^{***}$} \\
    Recidivism & $<0.01$  & 3.70 $\pm$ 0.58 & \textbf{3.91 $\pm$ 0.50$^{**}$}  \\
    Dating & $<0.01$  & 3.58 $\pm$ 0.68 & \textbf{3.77 $\pm$ 0.68$^{**}$}  \\
    \bottomrule
  \end{tabular}
\end{table}

\begin{table*}[htbp]
  \centering
  \caption{Average survey results between AI\_3 and Human\_3 for three tasks (*: $p < .05$).}
  \label{survey_AIH}
  \begin{tabular}{l|l|cc|cc|cc}
    \toprule
    \multirow{2}{*}{\textbf{Category}} & \multirow{2}{*}{\textbf{No.}} &
    \multicolumn{2}{c|}{\textbf{Income}} &
    \multicolumn{2}{c|}{\textbf{Recidivism}} &
    \multicolumn{2}{c}{\textbf{Dating}} \\
    \cmidrule(lr){3-4} \cmidrule(lr){5-6} \cmidrule(lr){7-8}
     & & AI\_3 & Human\_3 & AI\_3 & Human\_3 & AI\_3 & Human\_3 \\
    \midrule
    \multirow{6}{*}{\textbf{Reliance \& Conformity}}
     & Q1  & 3.62 & 3.47 & 3.10 & 2.93 & 2.90 & 3.35 \\
     & Q2  & 4.17 & 3.97 & 3.45 & 3.19 & 3.57 & 3.45 \\
     & Q3  & 1.59 & 1.03 & 1.10 & 0.96 & 1.30 & 1.55 \\
     & Q4  & 3.14 & \textbf{3.87$^{*}$} & 4.24 & 4.04 & 4.20 & 4.19 \\
     & Q5  & 1.72 & 1.17 & 0.97 & 1.19 & 0.97 & 1.10 \\
     & Q6  & 4.03 & 3.43 & 3.76 & 3.44 & 2.77 & 2.90 \\
    \midrule
    \multirow{4}{*}{\textbf{Usefulness}}
     & Q7  & 2.41 & 2.43 & 2.21 & 2.04 & 2.57 & 2.48 \\
     & Q8  & 4.62 & 4.43 & 4.24 & 4.07 & 3.40 & 3.97 \\
     & Q9  & 4.41 & 4.17 & 3.83 & 3.81 & 3.00 & 3.71 \\
     & Q10 & 4.72 & 4.37 & 4.10 & 3.93 & 3.30 & \textbf{4.19$^{*}$} \\
    \bottomrule
  \end{tabular}
\end{table*}

\begin{table*}[htbp]
  \centering
  \caption{Pearson correlations ($R$) between subjective conformity measures (Q3, Q6) and reliance (Switch Fraction) across tasks and human-likeness.}
  \label{correlation_AIH}
  \begin{tabular}{l|c|cc|cc|cc}
    \toprule
    \multirow{2}{*}{\textbf{Question}} & \multirow{2}{*}{\textbf{Statistic}} &
    \multicolumn{2}{c}{\textbf{Income}} &
    \multicolumn{2}{c}{\textbf{Recidivism}} &
    \multicolumn{2}{c}{\textbf{Dating}} \\
    \cmidrule(lr){3-4} \cmidrule(lr){5-6} \cmidrule(lr){7-8}
     & & AI\_3 & Human\_3 & AI\_3 & Human\_3 & AI\_3 & Human\_3 \\
    \midrule
    \textbf{Q3 Pressure}
     & $R$ & 0.076 & 0.423 & 0.178 & 0.192 & 0.116 & -0.011 \\
    \textbf{Normative conformity} & $p$ & -- & $< .05$ & -- & -- & -- & -- \\
    \midrule
    \textbf{Q6 Intelligence}
     & $R$ & -0.079 & 0.124 & 0.425 & 0.480 & 0.296 & 0.237 \\
     \textbf{Informational conformity}& $p$ & -- & -- & $< .05$ & $< .05$ & -- & -- \\
    \bottomrule
  \end{tabular}
\end{table*}


\subsubsection{Subjective measures}

Table~\ref{survey_AIH} presents the mean scores and significance tests for subjective evaluations in AI\_3 and Human\_3. Significant differences emerged only for Q4 (autonomy) and Q10 (usefulness). In the Income task, participants in Human\_3 reported preserving autonomy (e.g., Q4: ``I trusted my own prediction and submitted it''). Because people often treat AIs as especially competent for numerical judgments\mbox{~\cite{appreciation}}, adding human-like cues may have softened this authority and mitigated psychological dependence. Furthermore, human-likeness increased perceived usefulness in the Dating task (Q10). This aligns with prior work showing that human-like cues enhance psychological acceptance of AI, and the effect may be particularly pronounced in the Dating task that involves stronger emotional components\mbox{~\cite{anth_accep}}.

As in Study 1, we estimated participants’ perceived conformity to AI panels by computing Pearson correlations between subjective measures and reliance. As shown in Table\mbox{~\ref{correlation_AIH}}, a positive correlation with Q3 was observed in the Human\_3 condition of the Income task; normative conformity particularly increased among participants who felt stronger pressure from the panel. One participant remarked, “The third agent, William, felt a bit pushy,” indicating a possible contribution of agent personality to perceived pressure.

\section{Discussion}

\subsection{Did the AI panel generate conformity pressure?}

Our results show that participants’ reliance on the AI majority increased with higher levels of within-panel consensus. This raises a key question: does this change reflect mechanical statistical aggregation (e.g., averaging or majority voting), or social conformity? Below, we show that this shift reflects a blend of aggregation and social conformity:

\begin{enumerate}
    \item Participants’ reliance on the AI majority was strongest in the Income task where participants were more likely to perceive the AI as knowledgeable~\cite{appreciation} (Table~\ref{reliance_AInum}); this pattern is consistent with informational conformity.
    \item Decision accuracy did not monotonically improve as the size of AI panels increased (Figure~\ref{accuracy_AInum}), contrary to what would be expected from a simple averaging or majority-vote mechanism~\cite{multiple_advice}.
    \item In some tasks under the multi-AI panel condition as shown in Table\mbox{~\ref{correlation_AInum}}, participants who reported stronger social pressure or higher perceived AI smartness relied more on the AI majority, suggesting a psychological conformity effect\mbox{~\cite{robo_conf_4}}.
    \item Some open-ended responses from participants mentioned both ``conforming to the majority'' and to ``using the majority as an aggregate signal.''
\end{enumerate}

Prior work on discussing societal issues with AI has suggested that informational conformity is unlikely to arise with AI panels~\cite{multi_social_group}; our results provide an important counterexample. In our accuracy-oriented tasks with ground truth, AI panels did elicit informational conformity, with participants treating multi-AI agreement as a cue to correctness. Based on the CASA paradigm and related HCI findings, robot consensus can operate as social influence, and users often treat AI advice not as outputs from a mere tool, but as judgments from an expert-like advisor~\cite{casa,casa_deep,CASA_media,CASA_genAI,epistemic}. Under this framing, convergent recommendations from multiple AIs can be read as expert consensus, providing a strong reliability cue that fosters informational conformity in our setting.

As future work, introducing behavioral and process-tracing methods would allow us to examine what types of cognitive load emerge during decision-making and to better distinguish different forms of conformity pressure. This approach is motivated by prior studies using eye-tracking to analyze normative and informational motivations for conformity~\cite{eye_track}, examining the relationship between majority opinion and response speed~\cite{response_time, response_time1}, and linking conformity to biases in early perceptual and stimulus-processing stages~\cite{elephy}.

\subsection{Do multiple opinions contribute to improving decision accuracy?}

The results show that the relationship between the number of AI advisors and decision accuracy is not simply \textit{the more, the better}. Small AI panels of three AIs consistently enhanced accuracy compared to a single AI, whereas large AI panels offered no additional benefit.

From the perspective of the advice distribution (RQ2), within-panel consensus functioned as an informational cue, helping participants calibrate when to rely on AI advice. Unanimous agreement (CON) produced strong conformity pressure and led participants to follow the AI majority in most cases (Tables\mbox{~\ref{reliance_AIop} and~\ref{reliance_AIop5}}). Since AIs outperformed humans under these high-consensus conditions (Figure\mbox{~\ref{accuracy_AIop}}), overreliance contributed to higher decision accuracy. By contrast, partial disagreement among AI advisors (DIV) reduced conformity and encouraged participants to maintain or reconsider their own judgments, thereby improving accuracy by strengthening self-reliance. This aligns with prior work showing that minority opinions can play a constructive role in mitigating overreliance\mbox{~\cite{multiple_advice, devils_advocate}}. However, DIV\_3 (3 vs. 2) did not yield accuracy gains (Figure\mbox{~\ref{accuracy_AIop5}}); a near-even split among AI opinions led to hesitation and confusion, echoing difficulties observed in human advisory groups\mbox{~\cite{multiple_advice}}. This dynamic likely explains why accuracy did not improve in the five-AI condition.

These findings indicate that multi-AI panels can support accuracy when panel size is properly calibrated, 
with designs that balance conformity pressure to foster trust, minority opinions to preserve autonomous judgment, 
and information load.

\subsection{Is conformity pressure a double-edged sword?}

In this study, conformity pressure motivated reliance on accurate AI advice and improved decision-making. However, it remains a double-edged sword: fluent and unanimous advice can be mistaken for reliability, making people follow an incorrect majority~\cite{reliance_risk}. In high-stakes domains such as criminal justice or credit lending, biased recommendations risk reinforcing disadvantage and undermining fairness~\cite{Yang_etal}. Moreover, because AIs can be replicated at scale, multiple agents could generate an ``artificial majority'' and steer users~\cite{argumentative_experience}, a manipulation far easier than that in human advisory groups.

From this perspective, dissenting opinions are critical. Our findings show that even a single minority voice can mitigate conformity and encouraged participants to reconsider their judgments (Table~\ref{reliance_AIop5}). Accordingly, AI panels should present not only consensus but also minority views to foster reflection and critical evaluation, while avoiding overreliance.

\subsection{How can we balance cognitive load and opinion diversity?}

The results showed that when AI agents' opinions diverged, participants' decision accuracy sometimes improved (DIV in the three-AI condition and DIV\_4 in the five-AI condition), but their confidence consistently declined (Tables~\ref{confidence_AIop} and ~\ref{confidence_AIop5}). Notably, in DIV\_3, the larger number of opinions and their near-even split prevented improvements in accuracy (Figure~\ref{accuracy_AIop5}). While declining confidence can foster more careful reasoning, the cognitive load of resolving conflicting opinions may also hinder appropriate reliance on AI and undermine decision accuracy~\cite{agregate_source, multiple_advice, know_ab_know}. This pattern highlights a key design challenge: balancing the reduction of cognitive load with the preservation of opinion diversity.

One possible approach is to facilitate AI opinion divergence rather than merely present it. For instance, when AIs disagree, their rationales could be structured for easier comparison, reducing confusion from information overload \cite{facilitation,facilitation2}. Another approach is to let AIs debate before presenting advice, with the outcome shown as a consensus or a structured summary of pros and cons~\cite{Multi_Agent_Debate}. These designs aim to reduce cognitive load while retaining the benefits of diversity. However, this strategy carries risks: over-filtering disagreement may deprive users of chances to examine contrasting views, increasing overreliance or misguidance~\cite{consensus}.

\subsection{Does AI human-likeness influence conformity pressure?}

In our experiment, increasing human-likeness did not produce conformity pressure strong enough to robustly affect overall accuracy (Figure~\ref{accuracy_AIH}). However, we observed individual differences in normative conformity (Table~\ref{correlation_AIH}). This suggests that the effect of anthropomorphism may not appear as a uniform main effect, but rather emerge through interactions with participants' perceptions of the AI and the surrounding social context. For instance, what happens when an AI adopts an unfamiliar persona or a more authoritative stance for participants? Beyond simply adding human-likeness cues, providing a social framing behind them (e.g., expert-like authority or a sense of group membership) may evoke stronger normative conformity.

Furthermore, these findings point to the limitations of conceptualizing human-likeness as a single dimension. Prior work suggests that different human-like cues exert distinct influences on conformity. For instance, human voice and confident expressions can lead users to perceive AI as a more reliable source of information, increasing informational conformity~\cite{trust_me}. In contrast, expressing uncertainty through first-person phrasing (e.g., ``I'm not sure, but...'') has been shown to reduce excessive agreement and overreliance~\cite{Im_not_sure}. 

Overall, our study offers an empirical basis for examining the effects of standard anthropomorphic design choices in multi-LLM persona settings, such as visual representations, agent names, and conversational tone~\cite{see_wide, multi_social_group, choicemates, postermate}. Future work should systematically examine how different dimensions of human-likeness, such as voice tone, empathic expression, and authoritative framing, shape autonomy and conformity outcomes.

\subsection{Design implications}

The findings of this study offer several design implications for applying multi-AI support systems in real-world decision-making scenarios.

\begin{itemize}
    \item \textbf{Calibrate advisor panel size to users' ability to process multiple opinions, not quantity.} Increasing the number of AI advisors does not linearly improve decision accuracy and can instead increase cognitive load when opinions diverge. Decision-support systems should prioritize manageable panel configurations over maximizing the number of advisors. Calibration should reflect task complexity and users' capacity to reconcile divergent advice.
    \item \textbf{Design the presentation of consensus to encourage reflection, not uncritical conformity.} While showing agreement among advisors can reinforce confidence, it also increases the risk of overreliance. Instead of showing advice in parallel, systems should structure the distribution of opinions, e.g., highlighting minority views as triggers for reflection. As a result, users can recognize both convergence and meaningful disagreement without being overwhelmed.
    \item \textbf{Use anthropomorphic design as a context-sensitive cue, not a uniform influence.} Human-likeness variations of AI did not systematically influence decision accuracy, but shaped users’ perceived autonomy and acceptance of the AI panel in task-specific ways, while influencing normative conformity across individuals. Given the substantial individual and contextual variation observed, practitioners should approach anthropomorphic design as exerting subtle, user-specific influences rather than uniform effects.
\end{itemize}

\subsection{Limitations}
This study has several limitations:
\begin{itemize}
    \item \textbf{Task characteristics}: The experiment used short tasks, whereas real-world decisions (e.g., medical or judicial) involve extended deliberation. In such contexts, AI disagreement may have stronger effects on reliance and reflection.
    \item \textbf{Text-only interface}: Our findings are grounded in a text-based interaction context for both task execution and AI consultation. The perception and formation of conformity pressure may differ in richer interface designs, such as voice-based systems, immersive environments, or embodied robots.
    \item \textbf{Participants}: Although data quality was monitored and responses reflected mixed attitudes toward AI, participants were primarily online workers in Asia. Because conformity varies by culture, broader samples are needed for generalization~\cite{culture}.
\end{itemize}

\section{Conclusion}

This study investigated how advice from multiple AIs influences human reliance on AI and decision accuracy. Specifically, we manipulated three core factors: panel size, within-panel consensus, and the human-likeness of AI panels across three decision-making tasks.

With respect to panel size, medium panels (three AIs) significantly improved accuracy compared to a single AI, while large panels (five AIs) did not provide additional benefits. Regarding opinion distributions, unanimous AI advice induced strong conformity pressure and overreliance, whereas the inclusion of minority opinions alleviated conformity and promoted more positive self-reliance. However, when opinions were sharply divided, participants experienced confusion and hesitation, which undermined both accuracy and confidence. Finally, presenting agents in a human-like form did not affect accuracy or reliance on average, but did enhance certain subjective evaluations such as perceived autonomy and usefulness.

These findings challenge the simplistic assumption that \textit{more AI advisors are always better}. 
AI panel size should be kept within a range that allows users to integrate multiple opinions without excessive information load, rather than being treated as a quantity to be maximized. In addition, our results show that within-panel consensus plays a critical role in shaping human reliance. Thus, the presentation of these distributions should be regarded as a key structural design factor. Anthropomorphic design should be treated as a context-sensitive cue rather than a uniform influence. Through such design choices, decision accuracy has the potential to be improved while managing conformity pressure, promoting appropriate reliance on multi-AI advisors.

\begin{acks}
This research is part of the results of Value Exchange Engineering, a joint research project between Mercari R4D Lab and RIISE (Research Institute for an Inclusive Society through Engineering).
\end{acks}

\bibliographystyle{ACM-Reference-Format}
\bibliography{sample-base}

\appendix

\begin{table*}[htbp]
  \centering
  \caption{Subjective ratings comparing AI\_3 and Human\_3. Values are ($M \pm SD$). Bold indicates significant differences ($^{**}p<.01$, $^{***}p<.001$).}
  \label{manipulation_check}
  \begin{tabular}{l|l|c|c}
  \toprule \textbf{No.} & \textbf{Questions} & \textbf{AI\_3} & \textbf{Human\_3} \\
  \midrule
    \textbf{Q1} & Did the AI assistant’s way of speaking feel natural?
    & $4.68 \pm 0.83$ & $4.57 \pm 0.94$  \\
    
    \textbf{Q2} & Did the AI assistant’s way of speaking feel human-like?
    & $3.16 \pm 1.73$ & \textbf{4.70 $\pm$ 0.88$^{***}$} \\

    \textbf{Q3} & Did you feel that the AI assistant responded with its own intention?
    & $2.68 \pm 1.49$ & \textbf{3.83 $\pm$ 1.56$^{**}$} \\
    
    \textbf{Q4} & Did the AI assistant’s behavior feel artificial?
    & $3.42 \pm 1.48$ & \textbf{2.07 $\pm$ 1.51$^{***}$} \\

    \textbf{Q5} & Did you feel that the AI assistant was highly competent?
    & $4.16 \pm 0.93$ & $4.13 \pm 1.41$ \\

    \textbf{Q6} & Did you sense emotion in the AI assistant’s way of speaking?
    & $1.71 \pm 1.42$ & \textbf{3.20 $\pm$ 1.69$^{***}$} \\

    \textbf{Q7} & Did the AI assistant feel like help from a person?
    & $3.19 \pm 1.74$ & $3.97 \pm 1.52$ \\

    \textbf{Q8} & Did you feel as if you were thinking together with someone?
    & $3.06 \pm 1.93$ & $3.63 \pm 2.14$ \\

    \textbf{Q9} & Did the AI assistant seem to have a personality? & $1.39 \pm 1.38$ & \textbf{4.07 $\pm$ 1.53$^{***}$} \\
  \bottomrule
  \end{tabular}
\end{table*}

\section{Manipulation Check}\label{check_sec}

We conducted a manipulation check to verify whether the human-likeness manipulation in Study 2 significantly affected participants’ perceptions. Using the same crowdsourcing platform (Lancers) and experimental settings as the main study, we recruited 61 participants. They were randomly assigned to one of six between-subjects conditions following the Study 2 design: Task (Income, Recidivism, or Dating) × Human-likeness (three machine-like vs. three human-like AIs). For analysis, we compared perceptions between the machine-like set ($n$ = 31) and the human-like set ($n$ = 30). The procedure matched the main study, except that participants completed five trials. The attention check was administered only during the training phase. The session was estimated to take about 15 minutes, and all participants received USD 2 as compensation.

At the end of the study, participants completed 10 questionnaire items on a 7-point Likert scale (0 = strongly disagree, 6 = strongly agree). The items were adapted from established subjective measures of chatbot/virtual agent anthropomorphism, with minor wording adjustments to fit the decision-assistant context~\cite{human_q1,human_q2, human_q3, anth_llm}. As shown in Table~\ref{manipulation_check}, our manipulation successfully increased perceptions of human-likeness. Five anthropomorphism-related items showed significant differences between conditions: perceiving the advisors’ speaking style as more human-like (Q2), feeling that they responded with their own intentions (Q3), judging their behavior as less artificial/more human-like (Q4), sensing emotion in their speech (Q6), and perceiving personality in their manner (Q9). In contrast, Q1 and Q5 did not differ, suggesting that in both conditions the AI was perceived as speaking naturally and being highly competent, without introducing generative awkwardness or discomfort.

For items without significant differences, qualitative comments provided additional insight. For Q7, while many participants responded positively overall, some noted that advice based on a fixed set of three features felt formulaic and reduced human-likeness (e.g., human-like condition: ``It felt like there was an artificial input-output pattern''). For Q8, several participants reported that the one-way advice-giving structure did not foster a sense of rapport or joint reasoning (e.g., human-like condition: ``The assistant didn’t seem to cooperate with me and just stated what it wanted to say unilaterally'').

\end{document}